\documentclass[useAMS,usenatbib]{mn2e}

\usepackage{lscape,graphicx,pifont}
\usepackage{times}
\usepackage{xtab}
\usepackage{longtable}
\usepackage{natbib}

\newcommand{\Msun}{\mbox{\,M$_\odot$}}

\newcommand{\Rsun}{\mbox{\,R$_\odot$}}
\newcommand{\vunit}{\mbox{\,km\,s$^{-1}$}}
\newcommand{\mic}{\mbox{$\,\mu$m}}

\newcommand{\fion}[2]{[{#1}\,{\sc {#2}}]}
\newcommand{\ltsimeq}{\raisebox{-0.6ex}{$\,\stackrel
        {\raisebox{-.2ex}{$\textstyle <$}}{\sim}\,$}}
\newcommand{\gtsimeq}{\raisebox{-0.6ex}{$\,\stackrel
        {\raisebox{-.2ex}{$\textstyle >$}}{\sim}\,$}}

\newcommand{\sirtf}{\mbox{\it Spitzer Space Telescope}}
\newcommand{\spitzer}{\mbox{\it Spitzer}}
\newcommand{\HER}{\mbox{\it Herschel Space Observatory}}
\newcommand{\iso}{{\it ISO}}
\newcommand{\iras}{{\it IRAS}}

\newcommand{\RA}[3]{{#1}\,\,{#2}\,\,{#3}}
\newcommand{\dec}[3]{{#1}\,\,{#2}\,\,{#3}}

\title[A WISE view of novae -- I]{A WISE view of novae. I. The data}
\author[A. Evans et al.]{A. Evans$^{1}$\thanks{E-mail: a.evans@keele.ac.uk},
R. D. Gehrz$^{2}$, C. E. Woodward$^{2}$, L. A. Helton$^{3}$\\
$^{1}$Astrophysics Group, Keele University, Keele, Staffordshire, ST5 5BG, UK\\
$^{2}$Minnesota Institute for Astrophysics, School of Physics and Astronomy,
116 Church Street, S. E., University of Minnesota, Minneapolis, \\
Minnesota 55455, USA\\
$^{3}$ SOFIA Science Center, USRA, NASA Ames
Research Center, M.S. 232-12, Moffett Field, CA 94035, USA\\
}

\begin{document}

\date{Version of 20-05-2014}

\pagerange{\pageref{firstpage}--\pageref{lastpage}} \pubyear{2002}

\maketitle

\label{firstpage}


\begin{abstract}
We present the results of data-mining the Wide-field Infrared Survey Explorer
(WISE) archive for data on classical and recurrent novae. We find that the detections are
consistent with dust emission, line emission, emission by a stellar photosphere,
or a combination of these.
Of the 36 novae detected in one or more WISE bands, 16 are detected in all four;
thirty-one known novae are not detected by
WISE. We also searched for WISE data on post-WISE novae, to gain
information about nova progenitors.
In this first paper we consider only the WISE data.
In future papers we will provide a more detailed modelling of the WISE data, and
discuss WISE data on post-WISE novae -- including their variability -- and will
complement the WISE data with data from other IR surveys.
\end{abstract}


\section{Introduction}

Infrared (IR) observations of novae (both classical and recurrent) have been
carried out for over 40~years, starting with the IR photometry of FH~Ser by
\citeauthor{geisel} (1970; see \citealt{gehrz-CN2,BASI} for recent reviews).
Most IR observations are of individual objects, triggered once a nova is
reported. However, all-sky IR surveys provide a different perspective on the nova
phenomenon. These include the \iras\ \citep{iras} survey \citep[see][for a
discussion of novae and related objects as seen by \iras]{HG88,HG90,HG92,HG94}, AKARI
\citep{akari}, and 2MASS \citep[see][for a discussion of cataclysmic variables, 
including novae]{hoard} surveys
and, most recently, the Wide-Field Infrared Survey Explorer \citep[WISE;][]{WISE}.

\cite{hoard2} have reported the resuts of searching the WISE database for nova-like
variables. A preliminary report of the present work is given by \cite{evans-CT}.

\vspace{-4mm}

\section{The WISE survey}

\begin{figure}
\setlength{\unitlength}{1cm}
\begin{center}
\leavevmode
\begin{picture}(5.0,4.8)
\put(0.0,4.0){\includegraphics{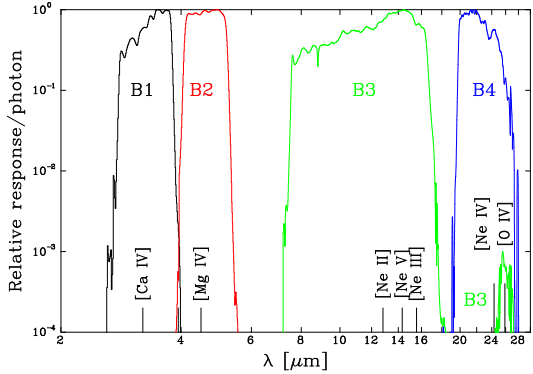}}
\end{picture}
\caption[]{Wavelength responses of WISE filters \citep{WISE}, together with the
locations of a sample of emission lines known to be strong in mature novae.
\label{wise-filters}} 
\end{center}
\end{figure}

WISE is an all-sky mid-infrared
mission \citep{WISE}, operating in wavebands centered at 3.4, 4.6, 12, and
22\mic. It commenced its survey on 2010 January 14, and completed its all-sky
survey on 2010 July 17; the 4-Band survey terminated on 2011 February 1, 2011 with the
exhaustion of hydrogen cryogen and after which WISE entered a 3-Band post-cryo survey phase.
Relevant calibration and other information for the WISE mission,
taken from \cite{WISE}, are given in Table~\ref{wise-char}.
The WISE bandpasses are shown in Fig.~\ref{wise-filters}.

\begin{table}
\caption{Characteristics of the WISE data \citep{WISE}. \label{wise-char}}  
\begin{tabular}{cccc}  \hline
WISE & Effective & Zero magnitude & Positional \\
Band & wavelength ($\mu$m) & flux (Jy) & accuracy (arcsec) \\ \hline
1 & 3.3 & 306.681 & $6.1$ \\
2 & 4.6 & 170.663 & $6.4$ \\ 
3 & 12  & 29.0448 & $6.5$ \\
4 & 22  & 8.2839  & $12.0$ \\
\hline\hline 
\end{tabular}
\end{table}

In this paper we present the data on classical and recurrent novae from the
WISE survey; in future papers we will discuss our modelling of the data, and
WISE data on post-WISE novae. While the thrust
of this work is the exploitation of the WISE survey in later papers we will supplement
the WISE data with data from other IR surveys as appropriate.

The prime purpose of the present paper is to present the WISE data, with a brief
discussion on a source-by source basis. For the present we discuss the data with
no allowance for colour correction \citep{WISE} -- although these are generally negligible
for ``photospheric'' temperatures -- or for reddening; we leave this to a more
detailed discussion in later papers.
Here we confine our discussion to those novae with detections in
at least three WISE bands, although in some cases detections in two WISE Bands are
sufficient to draw tentative conclusions.

\vspace{-6mm}

\section{WISE observations of novae}

In the course of its all-sky survey, WISE detected a number of classical and
recurrent novae. We data-mined the WISE catalogue
to identify coincident targets using source lists from \cite{BDE} for
pre-1989 novae, and \cite{DS1}, \cite{DS2}, \cite{DS3}, and \cite{SMARTS} for
later novae. We have confined our search to Galactic novae. J2000 positions were
taken from {\it SIMBAD}\footnote{http://simbad.u-strasbg.fr/simbad/},
\cite{SMARTS}, or \cite{BDE}.

The criterion for determining that WISE had indeed detected a nova was that the
WISE source position was within the WISE positional error of the nova
position; a visual check of the WISE 3.3\mic\ field was also performed.
Detections are given in Table~\ref{fluxes}, which includes outburst year. Also
given are $t_2$ and $t_3$, the times for the light curve (LC) to decline
by~ 2 and 3~magnitudes respectively, and the LC class;
the latter information is of the form X($t_3$), where X = S (smooth),
P (plateau), D (dust dip), C (cusp), O (oscillations), F (flat-topped),
J (jitter). The LC information is taken from \cite{strope} unless otherwise
noted in Table~\ref{fluxes}.

Table~\ref{fluxes} also includes the positions of the nova and associated WISE
source, the WISE fuxes (in mJy and {\bf Jy}), the difference $\Delta\theta$
(in arcsec) between the WISE position and that of the nova, and a crude classification 
of the WISE spectral energy distribution (SED). For the latter a ``$\cal{D}$'' indicates a
SED characteristic of dust emission, a ``$\cal{S}$'' indicates emission from a 
photosphere, a ``$\cal{L}$'' indicates likely line emission, and an $\cal{I}$
signifies inconclusive; in some cases a hybrid classification (e.g. $\cal{SD}$)
is possible. In many cases, the possibility of ``photospheric'' emission is deduced
on the basis of only two WISE Bands (usually~1 and~2) and the emission may be from
the secondary or an accretion disc; we do not discriminate between these two
possibilities at this stage. Where we have fitted blackbodies to the WISE data, the
uncertainty in the temperature is typically $\pm20$~K.

Novae not detected by WISE are listed in Table~\ref{non-det1}. These include objects with WISE
sources close to the nova position but not within the WISE positional uncertainties.

Classical novae (CNe) are well-known to go through a nebular (and in some cases,
coronal) phase, when emission lines dominate the spectrum. The location of some
of these lines in relation to the WISE bandpasses are shown in Fig.~\ref{wise-filters}
\citep[see][for a more complete list]{BASI}. The bandpasses encompass a
number of IR forbidden lines that are known to be strong in
mature (i.e., several hundreds days post-maximum) novae. Many
of these lines are neon lines that affect mainly WISE Bands~3 and~4, but there
are also coronal lines that affect WISE Bands~1 and~2. The ``red leak''
in Band~3 means that (for example) the \fion{O}{iv}~25.9\mic\ 
line, a strong coolant line of the plasma \citep{Woodward-Starrfield2011}
may, when present, contribute to the Band~3 flux.

\vspace{-4mm}

\section{Recurrent and suspected recurrent novae observed by WISE}

Recurrent novae (RNe) are a heterogeneous group of objects, in that the donor
star may be a main sequence dwarf or a red giant, that may or may not possess
a wind. Relatively few are known \citep[see][]{anupama}
compared with the number of known CNe, but apart from the nature of the donor
star the distinction between CNe and RNe is largely the selection effect that a
RN has been seen to erupt more than once. The timescale between RN eruptions is
$\ltsimeq$ a human lifetime, while that between CN eruptions is $\gtsimeq10^4$~years
\citep[see][]{gehrz-pasp}.
Since the sustained study of nove eruptions has a history of $\sim100$~years
it seems likely that inter-eruption timescales of $\ltsimeq100$~years and
$\gg100$~years are those that will be most evident at this stage of nova study:
it may be that there is a continuum of inter-eruption timescales that
has yet to become apparent \citep[cf.][]{darnley}. It is therefore clear
that, while all novae are recurrent, there are RNe masquerading as CNe
\citep{harrison,weight,darnley,pagnotta}.

Also there is evidence that RN eruptions occur on the surfaces of
massive white dwarfs (WDs) -- close to the Chandrasekhar limit -- and there is
great interest in determining whether the WD in some RN system is gaining mass,
possibly leading to deflagration as a Type~Ia supernova
supernova \citep[cf.][]{WhelanIben1973}.

Reviews of RNe, with emphasis on the RN RS~Oph, are given in \cite{RSOPH1} and
\cite{RSOPH2}, and a summary of RN properties (as of 2007) is given in
\cite{anupama}. \cite{strope} provide a discussion of RN light curves.

\vspace{-2mm}

\subsection{T CrB}
T~CrB is a recurrent nova that has undergone two known erutions, in 1866
and 1946 \citep{anupama}; its next eruption is overdue. The donor star
has spectral classification M3III \citep{anupama}.

The WISE SED for T~CrB is shown in Fig.~\ref{RNe}. The WISE data are
consistent with photospheric blackbody at $4\,300$~K \citep[cf. $\sim3\,500$~K
for a M3III star;][]{cox}. Within the WISE photometric uncertainties -- and
in contrast to RS~Oph \citep[see below and][]{evans-rs2} -- there is no
evidence for dust in the T~CrB system.

\vspace{-2mm}

\subsection{KT Eri}
KT~Eri erupted in 2009 and is a suspected RN \citep{darnley,pagnotta}. It is
exceptional for having a 
reasonably well-covered and extensive pre-eruption LC \citep{hounsell,WASP}. The rapid
light LC decline and the small amplitude suggest that it might be a RN
\citep{hounsell}. \cite{darnley} suspected that the donor star is evolved;
they give the near-IR colours as $R-I \simeq 0.6, J-H=0.42, H-K_s=0.05$;
\citeauthor{darnley} suggest
a classification of $\sim\mbox{G2III}$ and effective temperature $\sim5500$~K;
the reddening ($E(B-V)=0.08$) is too low to impact significantly on the IR colours.

\onecolumn

\setlongtables
\LTcapwidth=8.4in
\small
\begin{landscape}

{\tiny

\begin{table*}
\begin{center}
\caption{WISE fluxes. \label{fluxes}}  
\begin{tabular}{lrccllccccccc}  \hline
Nova & \multicolumn{1}{c}{Outburst} & $t_2, t_3$ & LC & \multicolumn{2}{c}{RA, Dec (J2000)}    & Date of first    & B1 (3.3\mic) & B2 (4.6\mic) & B3 (12\mic) &  B4 (22\mic) & $\Delta\theta$ & Classification/ \\
     & \multicolumn{1}{c}{year}    & (days) & class & \multicolumn{2}{c}{Upper value: WISE}  & WISE Observation & (mJy)       &   (mJy)      & (mJy)       &   (mJy)    & (arcsec) & Comment        \\
     &   &       & & \multicolumn{2}{c}{Lower value: GCVS} & YYYY-MM-DD       & ({\bf Jy})  &  ({\bf Jy}) & ({\bf Jy})   & ({\bf Jy})  &       &  \\ \hline


 V603 Aql & 1918 & 5, 12 & O(12) & \RA{18}{48}{54.64} & \dec{+00}{35}{03.43}  & 2010-04-01 & $14.63\pm0.40$ & $8.22\pm0.25$ & $<3.17$ & $<6.43$ & 0.51& $\cal{I}$ \\
          &  &   & & \RA{18}{48}{54.636} & \dec{+00}{35}{03.37}   &            &                 &                  &         &    && ---  \\

 V1229 Aql & 1970 & 18, 32 & P(32)  &\RA{19}{24}{44.56}  & \dec{+04}{14}{48.06}   & 2010-04-13 & $0.22\pm0.01$ & $0.12\pm0.02$ & $<0.36$ & $<2.48$ &  0.70& $\cal{I}$\\
           &   & &  & \RA{19}{24}{44.52}  & \dec{+04}{14}{48.7}   &            &                 &                  &         &    & & ---  \\

 V1370 Aql & 1982 &15, 28 & D(28) & \RA{19}{23}{21.32} & \dec{+02}{29}{25.43}  & 2010-04-12 & $0.33\pm0.02$ & $0.11\pm0.02$ & $<0.27$ & $<2.30$        & 2.69& $\cal{I}$\\
           &  &   & & \RA{19}{23}{21.10} & \dec{+02}{29}{26.1}   &            &                 &                  &         &    & & --- \\

 V1494 Aql & 1999 & 8, 16 & O(16) & \RA{19}{23}{05.29}   & \dec{+04}{57}{18.83}  & 2010-04-11 & $0.81\pm0.03$ & $0.45\pm0.02$ & $<0.31$ & $9.51\pm1.41$ & 0.27& $\cal{L}$ \\
           &   &  & &  \RA{19}{23}{05.30} & \dec{+04}{57}{19.1} &          &                 &                  &         &    &  & ---\\

 T Aur & 1891 & 80, 84 & D(84) & \RA{05}{31}{59.13}  &  \dec{+30}{26}{44.87} & 2010-03-08 & $1.16\pm0.03$ & $0.72\pm0.02$ & $0.40\pm0.15$ & $<3.36$ & 0.17& $\cal{I}$ \\
       &    &  & &  \RA{05}{31}{59.12} & \dec{+30}{26}{45.0}  &          &                 &                  &         &    &  & ---\\

 QZ Aur    & 1964 &--, 23--30 & -- &  \RA{05}{28}{34.08}  & \dec{+33}{18}{21.65} & 2010-03-08 & $0.33\pm0.01$ & $0.17\pm0.02$ & $<0.28$ & $<2.24$ & 0.15& $\cal{I}$ \\
           &  &   & & \RA{05}{28}{34.08} & \dec{+33}{18}{21.8}  &          &                 &                  &         &    &  & $t_3$ from [1] \\

T Boo  & 1860 & --, -- & -- & \RA{14}{14}{1.97} & \dec{+19}{04}{3.79} & 2010-01-14   & $0.50\pm0.01$ & $0.30\pm0.01$    & $<0.23$     & $2.05\pm0.81$ & 1.74& $\cal{I}$ \\
       &   &  & & \RA{14}{14}{06\,.\,~~} & \dec{+19}{03}{54~~} &    &  &    &    & &  & ---\\

 V705 Cas & 1993 & 33, 67 & D(67)  & \RA{23}{41}{47.18}  & \dec{+57}{30}{59.65} &2010-01-12 & $0.80\pm0.02$ & $0.49\pm0.02$ & $1.15\pm0.17$ & $13.08\pm1.06$ & 0.18& ${\cal L}$ \\
          &  &  & & \RA{23}{41}{47.19}  & \dec{+57}{30}{59.5} &  &                 &                  &         &    & & --- \\

 V723 Cas & 1995 &263, 299& J(299) & \RA{01}{05}{05.34}  & \dec{+54}{00}{40.35}  & 2010-01-22 & $1.34\pm0.03$ & $0.81\pm0.02$ & $0.55\pm0.090$ & $<1.37$ & 0.19& ${\cal SL}$ \\
           &  &  & & \RA{01}{05}{05.36} & \dec{+54}{00}{40.3}  &  &                 &                  &         &    &  & ---\\

 V1065 Cen & 2007&11, 26 & -- &\RA{11}{43}{10.30} & \dec{--58}{04}{04.85} & 2010-01-18 & $0.58\pm0.05$& $0.44\pm0.03$ & $12.75{\pm}0.28$ & $29.53\pm1.36$ & 0.56& ${\cal DL}$\\
            & &  & &  \RA{11}{43}{10.33} & \dec{--58}{04}{04.3}  &  &  &  &  &  &  & $t_2,t_3$ from [3] \\

T CrB  & R1946   &4, 6&  S(6) & \RA{15}{59}{30.14} & \dec{+25}{55}{12.68} & 2010-02-05   & {\bf 4.70}$\pm${\bf 0.78} & {\bf 2.70}$\pm${\bf 0.19} & {\bf 0.50}$\pm${\bf 0.01} & {\bf 0.17}$\pm${\bf 0.01} & 0.20& $\cal{S}$; [2] \\
       &   &  & & \RA{15}{59}{30.16}& \dec{+25}{55}{12.59}  &    &  &  &  &  & & --- \\

AR Cir    & 1906 & --, -- & -- &  \RA{14}{48}{09.51} & \dec{--60}{00}{27.48} & 2010-02-17 & $15.17\pm0.38$ & $7.92\pm0.18$ & $<2.68$ & $8.13\pm1.92$ & 0.27& $\cal{SI}$ \\
          &  & &   & \RA{14}{48}{09.53} & \dec{--60}{00}{27.5} & & & & & && Symbiotic?\\

DZ Cru & 2003 &--, --& -- & \RA{12}{23}{16.08} & \dec{--60}{22}{34.34}& 2010-01-26   & $53.3\pm0.01$ & $156\pm1$ & $1075\pm12.9$ & 
                                                         $963\pm11.5$ & 1.38& $\cal{D}$ \\ 
       &   &  & & \RA{12}{23}{16.2} & \dec{--60}{22}{34}&             &       &   &  &  & & --- \\

V476 Cyg  & 1920 & --, 16.5 & -- & \RA{19}{58}{24.46} & \dec{+53}{37}{07.56}  & 2010-05-20 & $0.26\pm0.01$ & $0.15\pm0.01$ & $<0.14$   & $<1.23$ & 0.06& $\cal{I}$ \\ 
         & &   &   & \RA{19}{58}{24.46} & \dec{+53}{37}{07.5} & & & & & & & $t_3$ from [1]\\

V1974 Cyg &1992 &19, 43  & P(43) &\RA{20}{30}{31.60}  & \dec{+52}{37}{51.85}  & 2010-05-29 & $0.25\pm0.01$ & $0.15\pm0.01$ & $0.40\pm0.07$ & $2.59\pm0.65$ & 0.57& $\cal{L}$ \\
          &  &  & & \RA{20}{30}{31.61}  & \dec{+52}{37}{51.3}  &  &  &  &  &  & & --- \\

V2361 Cyg & 2005 & --, -- & & \RA{20}{09}{18.78}  & \dec{+39}{48}{51.74} & 2010-05-31 & $0.54\pm0.02$    & $0.11\pm0.01$ & $<0.63$ & $6.37\pm1.14$ & 2.42& $\cal{I}$ \\
         &   &  & & \RA{20}{09}{19.05} & \dec{+39}{48}{52.9} &  && & & & & ---\\

V2362 Cyg & 2006 & 9, 246 & C(246) &\RA{21}{11}{31.88}  & \dec{+44}{48}{03.16} & 2010-05-31 & $1.51\pm0.05$ & $0.78\pm0.02$ & $1.18\pm0.11$ & $13.60\pm0.90$ & 4.76& $\cal{I}$  \\
          &&   &   & \RA{21}{11}{32.346} & \dec{+44}{48}{03.66} &  && & & && --- \\

V2467 Cyg & 2007 & 8, 20 & O(20) &\RA{20}{28}{12.48}  & \dec{+41}{48}{36.52} & 2010-05-17 & $2.28\pm0.11$ & $3.58\pm0.15$ & $22.22\pm1.37$ & $226.1\pm9.4$ & 0.38& $\cal{LD}$?  \\
          &&   &   & \RA{20}{28}{12.52} & \dec{+41}{48}{36.5} &  && & & && --- \\
          
\hline\hline 
\end{tabular}
\end{center}
\end{table*}

}

\setcounter{table}{1}

{\tiny

\begin{table*}
\begin{center}
\caption{WISE fluxes (continued)}  
\begin{tabular}{lrccllccccccc}  \hline

Nova & \multicolumn{1}{c}{Outburst} & $t_2, t_3$ & LC & \multicolumn{2}{c}{RA, Dec (J2000)}    & Date of first    & B1 (3.3\mic) & B2 (4.6\mic) & B3 (12\mic) &  B4 (22\mic) &
$\Delta\theta$ & Classification/ \\
     & \multicolumn{1}{c}{year}  & (days)  & class  & \multicolumn{2}{c}{Upper value: WISE}  & WISE Observation & (mJy)       &   (mJy)      & (mJy)       &              & (arcsec) & Comment        \\
     &    &   &   &  \multicolumn{2}{c}{Lower value: GCVS} & YYYY-MM-DD       & ({\bf Jy})  &  ({\bf Jy}) & ({\bf Jy})   & ({\bf Jy})  &         \\ \hline
	  
HR Del & 1967 & 167, 231 & J(231) & \RA{20}{42}{20.35} & \dec{+19}{09}{39.22} & 2010-05-09   & $5.61\pm0.12$ & $3.63\pm0.07$ & $2.20\pm0.14$ & $33.35\pm1.47$ & 0.09& $\cal{L}$ \\
    &   &    &  & \RA{20}{42}{20.34} & \dec{+19}{09}{39.31}  &   &  &  &  &  & & ---  \\

KT Eri & R?2009  & --, --& -- & \RA{04}{47}{54.20}  & \dec{--10}{10}{42.74}  &2010-02-20   & $27.64\pm0.59$ & $11.84\pm0.22$  & $1.96\pm0.31$ & $10.78\pm0.97$ & 0.37&
$\cal{SI}$; [4]\\
    &   &  & & \RA{04}{47}{54.21}  & \dec{--10}{10}{43.1}  &    &   &    &  &   & & $t_2$ from [5,6] \\

DM Gem    & 1903 &--, 22 & -- & \RA{06}{44}{11.55}  & \dec{+29}{56}{43.71} & 2010-03-22 & $0.75\pm0.02$ & $0.38\pm0.02$ & $<0.40$   & $<3508$ & 1.97& $\cal{I}$ \\
         & &   &    & \RA{06}{44}{12.05} & \dec{+29}{56}{41.9}  &   & & & & & &  $t_3$ from [7] \\

DN Gem    & 1912 & 16, 35 & P(35)  & \RA{06}{54}{54.35}  & \dec{+32}{08}{27.52} & 2010-03-22 & $0.22\pm0.03$ & $0.13\pm0.02$ & $<0.93$    & $<2.03$ & 0.48& $\cal{I}$ \\
          &&    &   & \RA{06}{54}{54.35} & \dec{+32}{08}{28.00}  &   & & & & & & $t_3$ from [1]\\

DQ Her & 1934 & 76, 100 & D(100) & \RA{18}{07}{30.26}  & \dec{+45}{51}{32.55}  & 2010-03-22  & $1.94\pm0.03$ & $1.24\pm0.03$  & $0.34\pm0.10$ & $<2.69$ & 0.07& $\cal{S}$  \\
       &  & &  & \RA{18}{07}{30.25}  & \dec{+45}{51}{32.6}  &     & &    &   &   & & --- \\

V533 Her & 1963 & 30, 43 & S(43) & \RA{18}{14}{20.49}  & \dec{+41}{51}{22.39} & 2010-03-23 & $0.55\pm0.01$ & $0.34\pm0.01$ & $<0.27$ & $<1.47$ & 0.30& $\cal{I}$ \\
         &  &  &  & \RA{18}{14}{20.48} & \dec{+41}{51}{22.1} & & & & &  & & --- \\

RS Oph & R2006& 7, 14 & P(14) & \RA{17}{50}{13.16} & \dec{--06}{42}{28.52}  & 2010-03-17  & {\bf 1.04}$\pm${\bf 0.12} & {\bf 0.72}$\pm${\bf 0.03} & $252\pm3$ & $153\pm4$ & 0.5& $\cal{SD}$; [3] \\
    &   & & & \RA{17}{50}{13.20}  & \dec{--06}{42}{28.48}  &   &  &  &  & & &  --- \\

GK Per & 1901 & 6, 13& O(13) & \RA{03}{31}{12.00} & \dec{+43}{54}{15.24}  & 2010-02-12    & $36.40\pm0.77$ & $22.46\pm0.46$ & $4.72\pm0.17$ & $2.51\pm0.86$ & 3.12& $\cal{S}$ \\
       &   &   & & \RA{03}{31}{11.82} & \dec{+43}{54}{16.8}  &     &   &   &   &   & & ---  \\

RR Pic & 1925 & 73, 122 &J(122)  & \RA{06}{35}{36.08}   & \dec{--62}{38}{24.31}    & 2010-04-10    & $4.47\pm0.10$ & $3.09\pm0.06$  & $0.94\pm0.061$ & $<1.33$ & 0.16 & $\cal{SI}$ \\
       &  &  &  & \RA{06}{35}{36.06}  & \dec{--62}{38}{24.37}  &     &  &   &  & &  & ---\\

T Pyx & R2011 & 32, 62 &P(62)  & \RA{09}{04}{41.51}   & \dec{--32}{22}{47.61}    & 2010--05--14    & $0.91\pm0.02$ & $0.67\pm0.02$  & $<0.28$ & $<4.52$ & 0.15 & ${\cal I}$; [3]\\
       &  &  &  & \RA{09}{04}{41.50}  & \dec{--32}{22}{47.5}  &     &  &   &  & & &---\\


V3890 Sgr & R1990 & 6, 14 & -- & \RA{18}{30}{43.28} & \dec{--24}{01}{08.97}  & 2010-03-26& $216\pm5$ & $119\pm2$ & $33.2\pm0.6$ & $12.93\pm1.42$ & 0.08& $\cal{S}$; [3] \\
          &  &  &   & \RA{18}{30}{43.28} & \dec{--24}{01}{08.9}  & &  &  & &  & & --- \\

U Sco  & R2010 & 1, 3 &P(3)  & \RA{16}{22}{30.80}  & \dec{--17}{52}{43.28}  &2010-02-24   & $1.58\pm0.04$ & $0.77\pm0.02$   & $0.78\pm0.15$ & $<2.89$ & 0.08& $\cal{L}$?; [3] \\
       &  &  &   & \RA{16}{22}{30.80}  & \dec{--17}{52}{43.2}  &   &  &   &  & & & --- \\

V745 Sco &  R1989 & --, -- & -- & \RA{17}{55}{22.23}  & \dec{--33}{14}{58.62} & 2010-03-18 & $330\pm011$ & $248\pm5$ & $131\pm2$ & $64.36\pm2.08$ & 0.21 & $\cal{SD}$; [3] \\
         & &  &     & \RA{17}{55}{22.27} & \dec{--33}{14}{58.5}   & &    & & & & & ---\\

V1186 Sco & 2004 & 12, 62 & J(62) & \RA{17}{12}{51.26} & \dec{--30}{56}{37.33} &2010-03-09 & $1.91\pm0.10$ & $2.10\pm0.08$ & $7.16\pm0.26$ & $4.07\pm1.19$ & 0.79& $\cal{L}$ \\
          & &  & & \RA{17}{12}{51.21} & \dec{--30}{56}{37.2} &  &  &  &  &  & &  --- \\

V1280 Sco & 2007 & --, 34& -- & \RA{16}{57}{41.20} & \dec{--32}{20}{36.40} &2010-03-04 & {\bf 2.53}$\pm${\bf 0.21} & {\bf 7.58}$\pm${\bf 0.37} & {\bf 24.29}$\pm${\bf 2.48} & {\bf 13.95}$\pm${\bf 0.15} & 2.74& $\cal{D}$ \\
          & &  & & \RA{16}{57}{40.91} & \dec{--32}{20}{36.4} &  &  &  &  &  & & $t_3$ from [2] \\
	  
EU Sct & R?1949 & --, --& -- & \RA{18}{56}{13.23}  & \dec{--4}{12}{32.26}  & 2010-04-02  & $22.78\pm0.44$ & $11.62\pm0.23$ & $2.68\pm0.23$ & $<2.33$ & 0.76& $\cal{S}$; [4] \\
       &   & &  & \RA{18}{56}{13.12}  & \dec{--4}{12}{32.3}   &    &  &  &   &  &  & --- \\
       
CT Ser & 1948 & --, $>100$? & -- & \RA{15}{45}{39.10} & \dec{+14}{22}{31.77} & 2010-02-05   & $0.18\pm0.01$ & $0.11\pm0.01$    & $<0.29$     & $<2.64$  & 0.07& $\cal{I}$ \\
       &  & &   & \RA{15}{45}{39.08} & \dec{+14}{22}{31.8} &    &  &    &     &   & &  $t_3$ from [1]\\

V382 Vel &1999 & 6, 13 & S(13) & \RA{10}{44}{48.45}  & \dec{--52}{25}{30.79}  & 2010-01-07 & $0.42\pm0.02$ & $0.29\pm0.01$ & $4.33\pm0.13$ & $6.41\pm0.72$ & 0.49& $\cal{L}$ \\
         &  & &  & \RA{10}{44}{48.39}  & \dec{--52}{25}{30.7}  &  &  &   &   &   & & ---  \\

\hline\hline 
\multicolumn{10}{l}{[1] \cite{OBBode}; [2] \cite{v1065cen}; [3] Recurrent nova; 
[4] Potential recurrent; [5] \cite{hounsell}; [6] \cite{WASP}; [7] \cite{HG90}}&&\\
\end{tabular}
\end{center}
\end{table*}

}

\end{landscape}

\twocolumn

The WISE SED is shown in Fig.~\ref{RNe}. All we can say is that the Band 1--3 data
are just consistent with the Rayleigh-Jeans tail of a blackbody, but the temperature
is poorly constrained by the WISE data alone; however there remains a large excess
in Band~4. The WISE data do not discriminate between line and dust emission.

\subsection{RS Oph}
The best-studied of the RNe, RS~Oph has undergone five known and two suspected
eruptions \citep{anupama}. The donor star is an early (M0/M2) giant with a $^{12}$C/$^{13}$C
ratio of $16\pm3$, an underabundance of C and an overabundance of N relative to solar values
\citep{pavlenko-aa,pavlenko-mn}.
\citeauthor{pavlenko-aa} determined an effective temperature of $4\,100\pm100$~K for the
donor star. Silicate dust -- seemingly oblivious to the 2006 eruption
-- was prominent in a spectrum obtained with the 
Infrared Spectrograph \citep[IRS;][]{houck} on the \sirtf\
\citep{spitzer,gehrz-spitzer} in 2007 \citep{evans-rs2,woodward-rs}.

\begin{figure*}
\setlength{\unitlength}{1cm}
\begin{center}
\leavevmode
\begin{picture}(5.0,22)
\put(0.0,4.0){\includegraphics{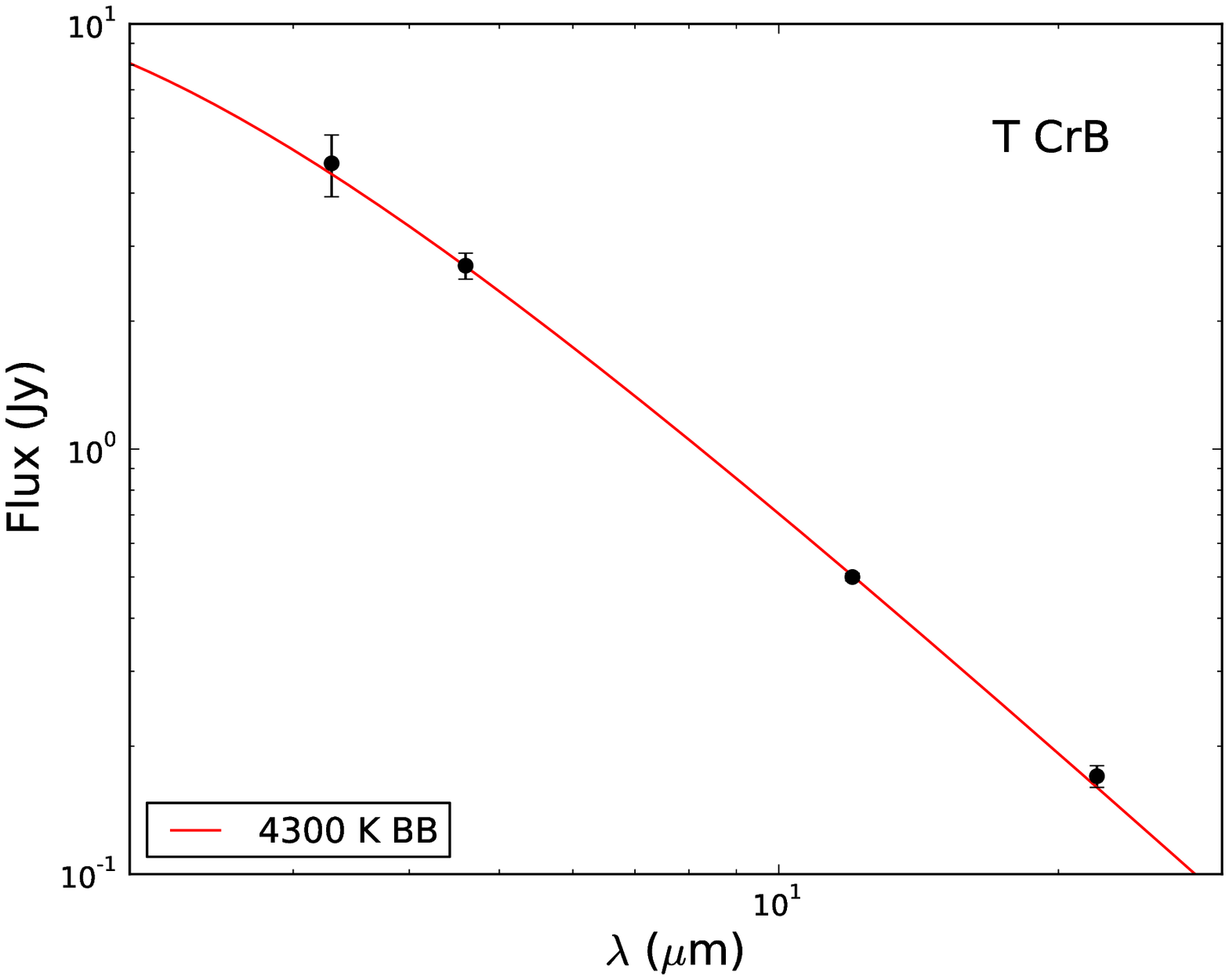}}
\put(0.0,4.0){\includegraphics{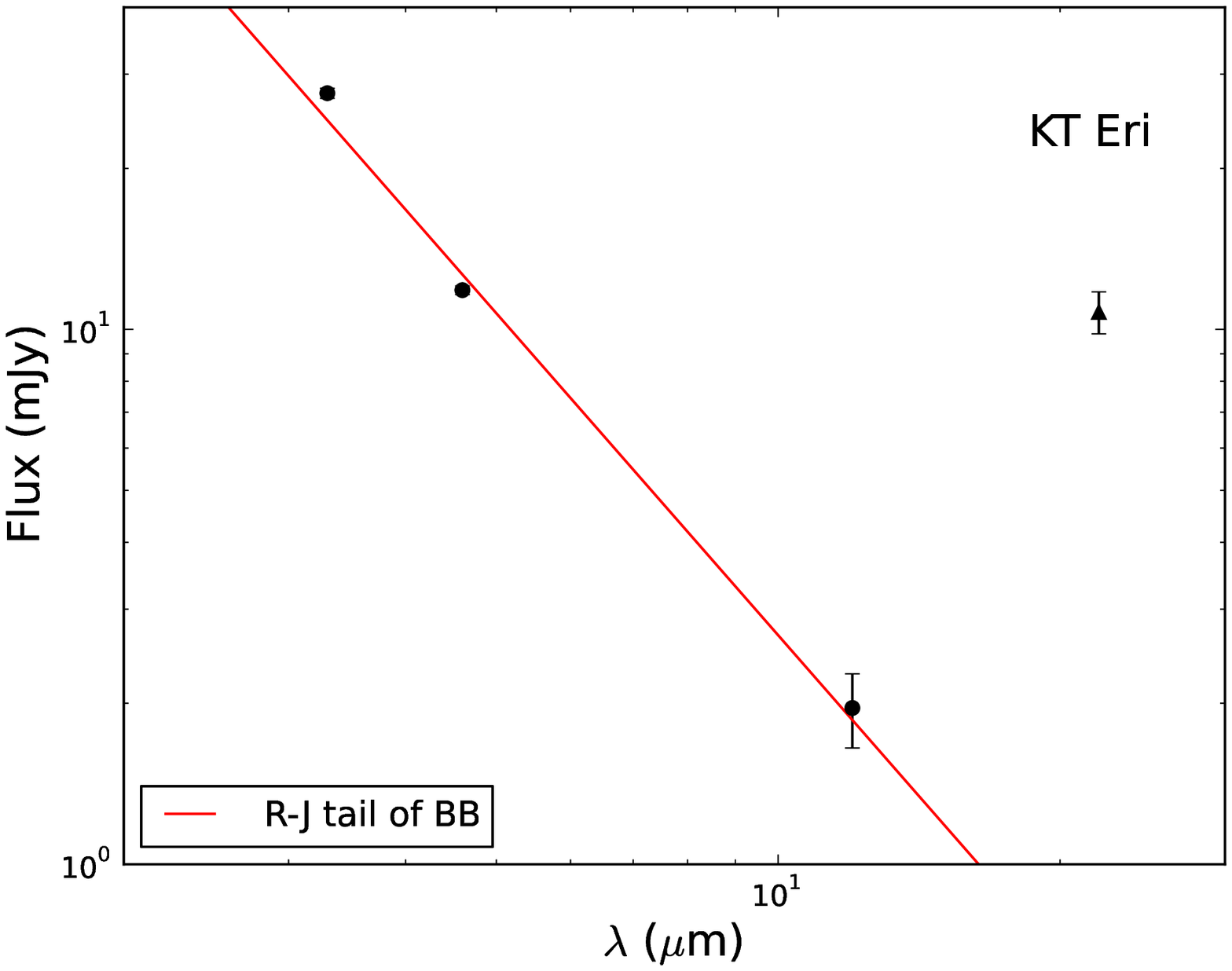}}
\put(0.0,4.0){\includegraphics{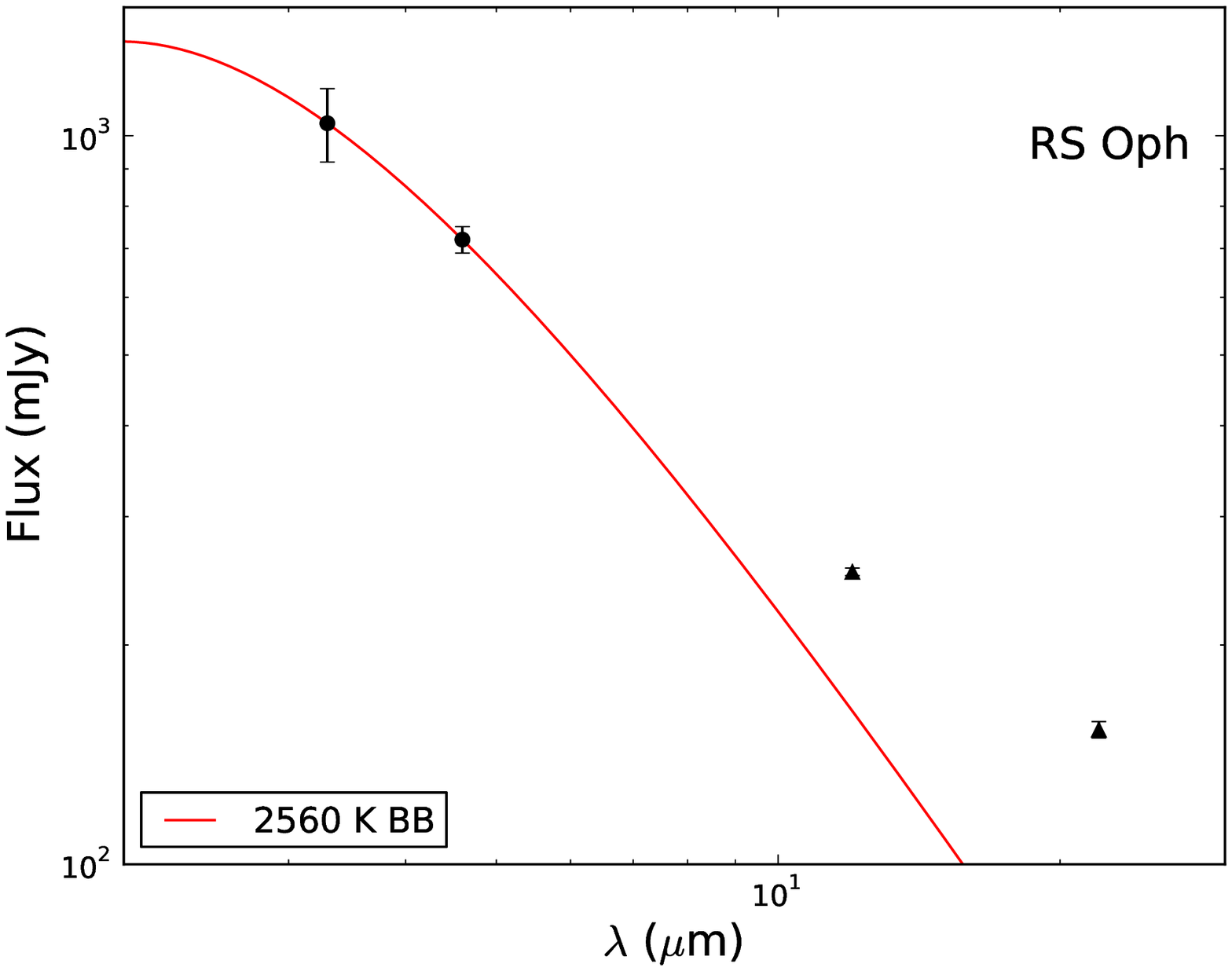}}
\put(0.0,4.0){\includegraphics{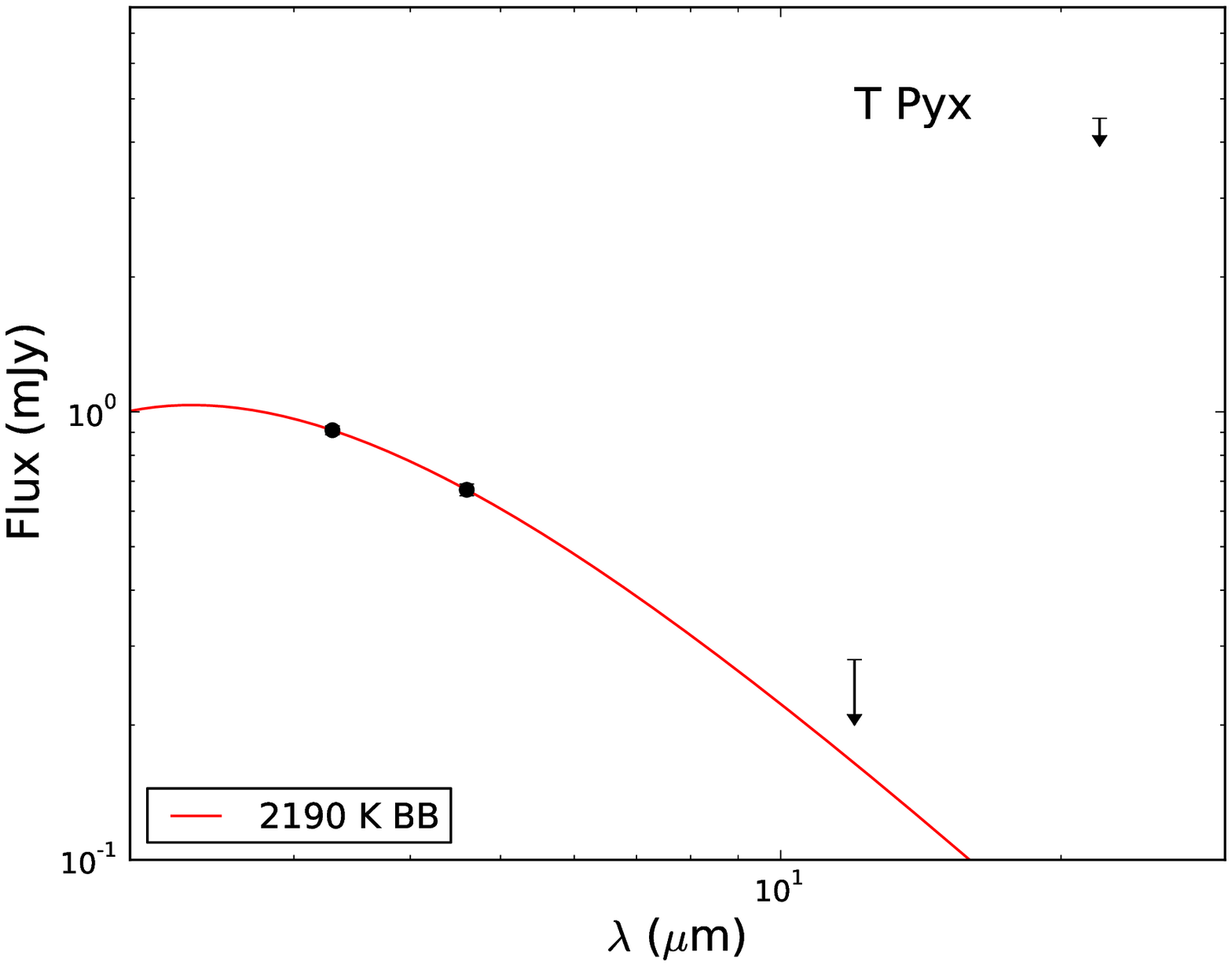}}
\put(0.0,4.0){\includegraphics{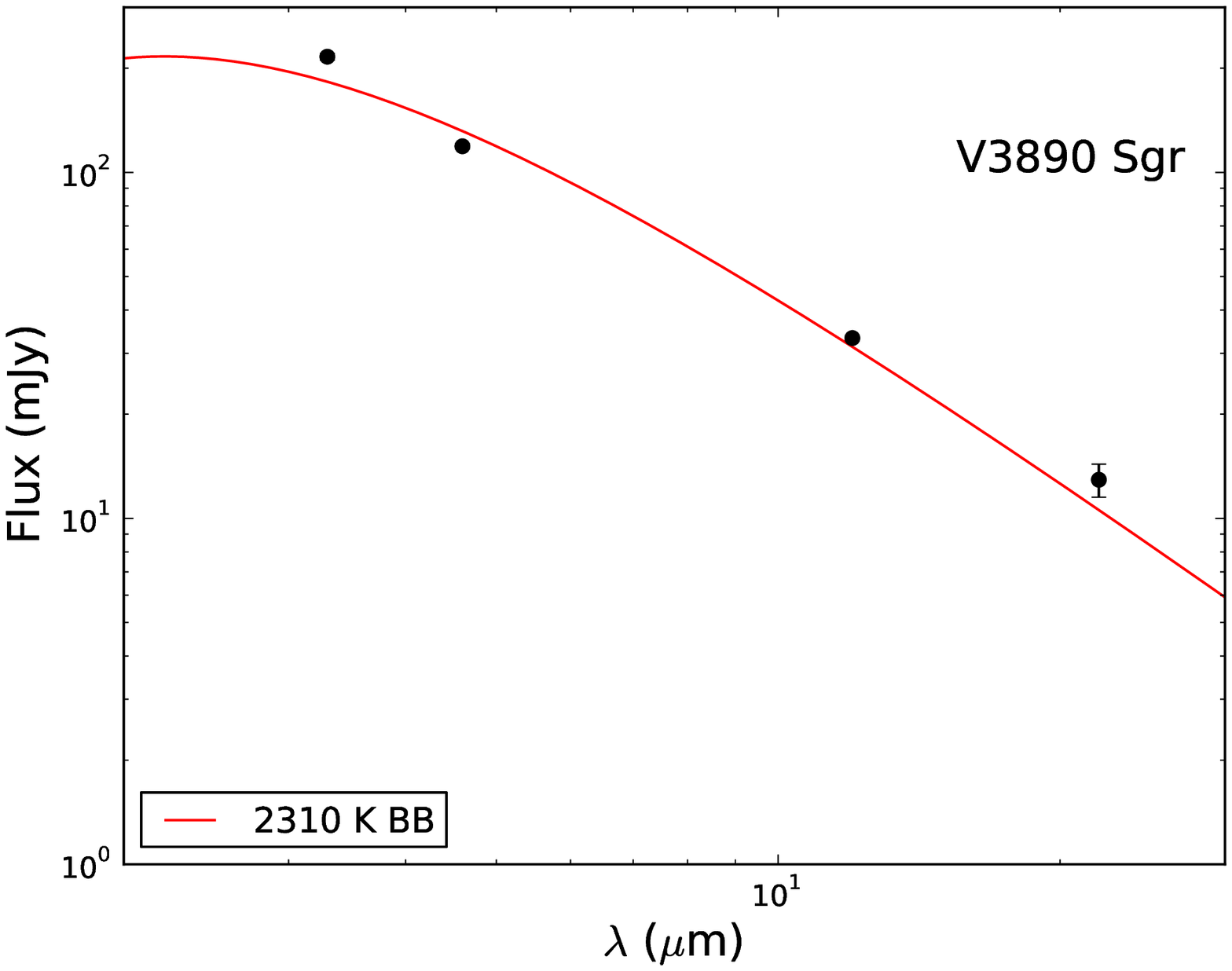}}
\put(0.0,4.0){\includegraphics{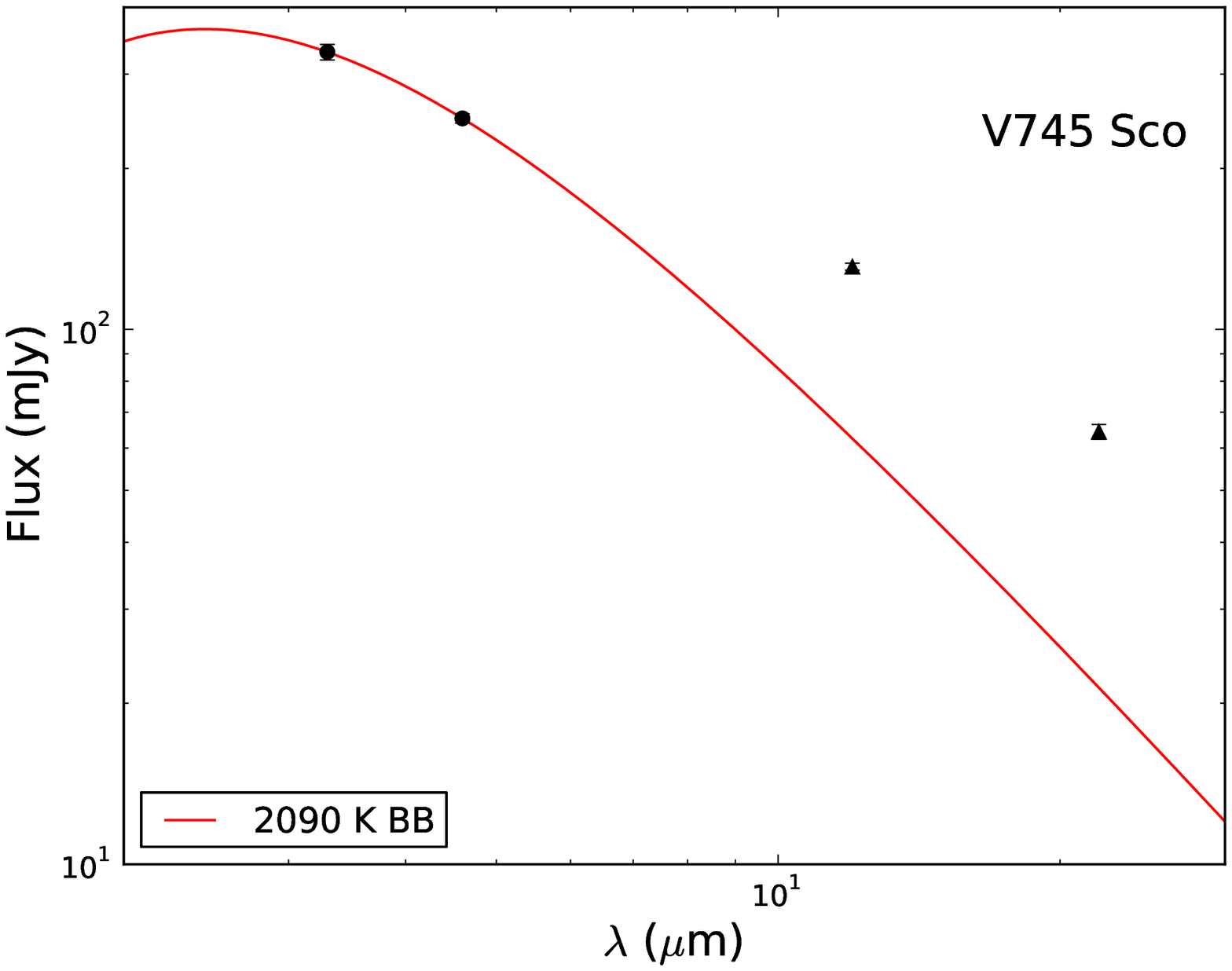}}
\put(0.0,4.0){\includegraphics{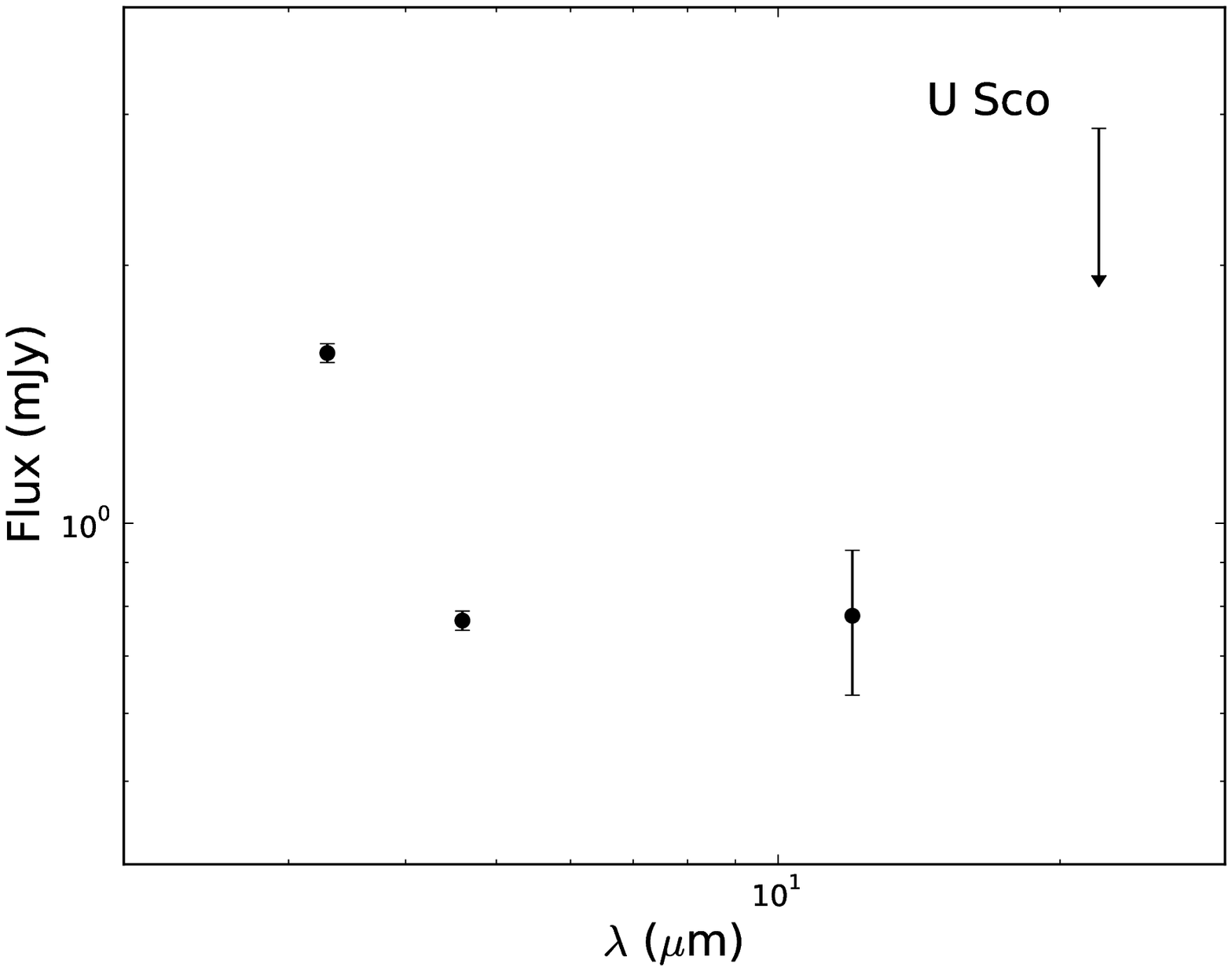}}
\put(0.0,4.0){\includegraphics{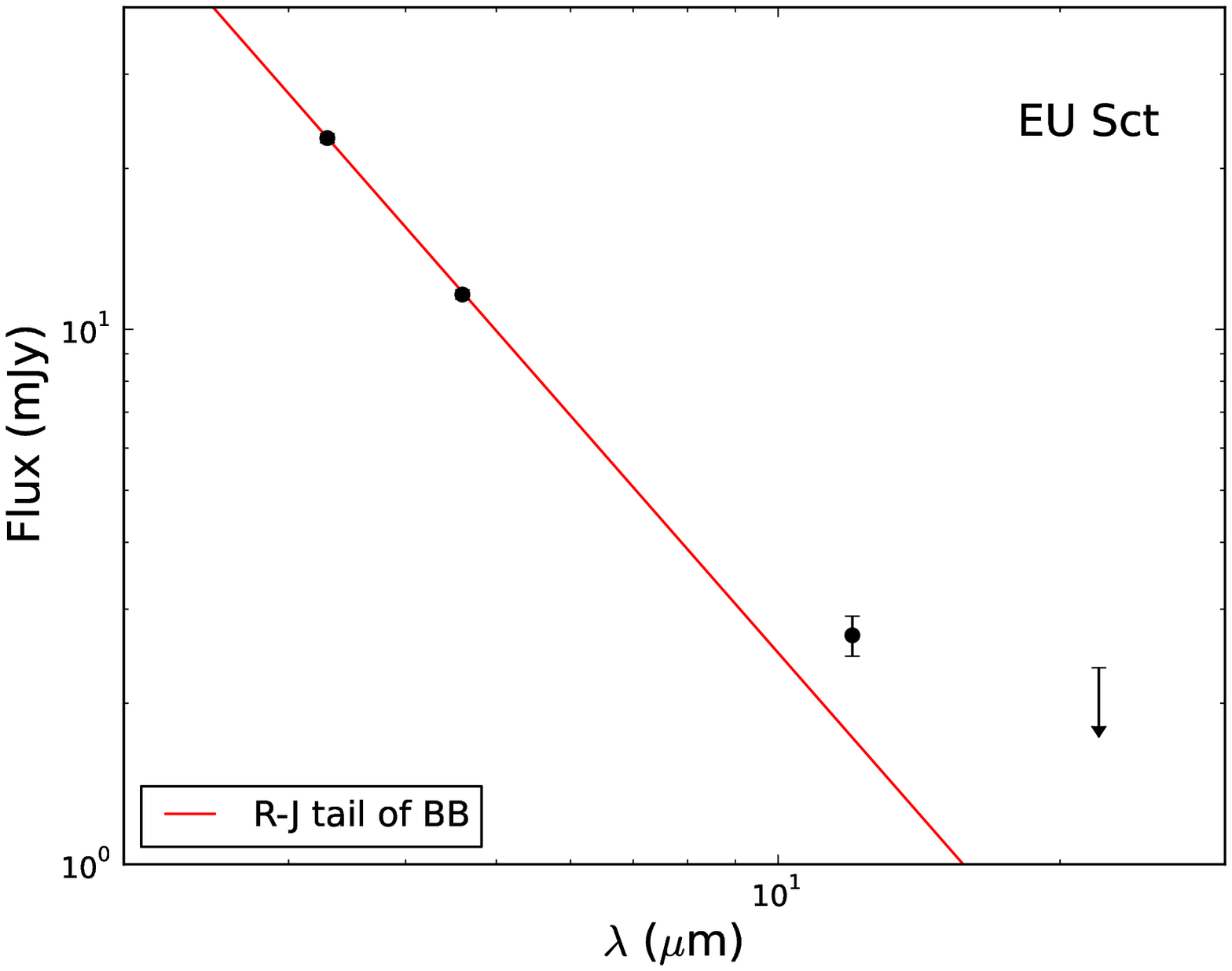}}
                 \end{picture}
\caption[]{Recurrent novae and suspected recurrent novae in the WISE database.
Top left: TCrB.
Top right: KT Eri.
Upper left middle: RS Oph.
Upper right middle: T Pyx.
Lower left middle: V3890 Sgr.
Lower right middle: V745 Sco.
Bottom left: U Sco.
Bottom right: EU Sct.
Errors are smaller than plotted points if not shown.
\label{RNe}} 
\end{center}
\end{figure*}

The WISE SED is shown in Fig.~\ref{RNe}. The Band 1--2 data seem consistent with a 2\,560~K
blackbody, leaving an excess in Bands 3--4; we identify this with the dust
in the RS~Oph system, although the \fion{O}{iv} fine structure line
-- commonly seen in CNe \citep{helton} and indeed present in the IR spectrum of
RS~Oph \citep{evans-rs1} -- may contribute in Band~4. 
We defer a discussion of this point to detailed modelling.

\subsection{T~Pyx}
T~Pyx is a well-known recurrent nova which underwent its sixth known eruption in
2011. The nature of the donor star is uncertain.
Observations with the \HER\ during the 2011 eruption
revealed a weak far-IR excess \citep{evans-tpyx}, attributed to an IR echo from pre-existing
dust.

T~Pyx is detected in WISE Bands 1~and 2~only (see Fig.~\ref{RNe}). The Bands 1 and 2
data are consistent with a 2\,190~K BB but in view of the limited nature of the data
this is not conclusive.

\subsection{V3890 Sgr}

V3890~Sgr is a RN that underwent eruptions in 1962 and 1990.
The secondary star is classed as M5III \citep{anupama}. The WISE SED is consistent
with a blackbody photosphere at 2\,310~K (Fig.~\ref{RNe}). This is
somewhat cooler than the effective temperature of a M5III \citep[3\,170~K][]{cox}.
There is no evidence for either line emission or a dust excess.

\subsection{V745 Sco}

V745~Sco underwent RN eruptions in 1937 and 1989;
the secondary star is classed as M6III \citep{anupama}, corresponding to an
effective temperature of 3\,250~K.
The WISE SED seems (like that of RS~Oph) to be consistent with a 
blackbody at 2\,090~K, together with a dust excess (see Fig.~\ref{RNe}).

\subsection{U Sco}
U~Sco is a well-known eclipsing RNe with eruptions in 1863, 1906, 1936, 1979, 1987,
1999 and in 2010; a possible eruption in 1917 has been identified \citep{anupama}.
The secondary is constrained to be between a F3 and a G sub-giant \citep{mason1};
\cite{anupama} gives K2IV (effective temperature 4\,620~K).

The WISE SED is shown in Fig.~\ref{RNe}. The interpretation of the WISE SED is 
constrained by the fact that the $J$-band flux is $\sim0.3$~mJy \citep{darnley}:
there is no blackbody at the temperature of a K2IV subgiant that is consistent
with the WISE and $J$-band fluxes. It is likely therefore that the IR SED
is due to the presence of emission lines.

\subsection{EU Sct}

EU Sct had an eruption in 1949 and was identified as a potential RN by
\cite{weight} on the basis of its $J\!H\!K$ colours \citep[see
however][]{pagnotta}; the near-IR colours are
$J-H=1.08, H-K_s=0.38$ by \cite{darnley}; \citeauthor{darnley} and \cite{weight}
give the reddening as $E(B-V)=0.84$, so that the dereddened colours correspond
to an early M giant classification. 

The WISE data shown in Fig.~\ref{RNe}. The Band~1--2 data are consistent only with the 
Rayleigh-Jeans tail of a blackbody so the temperature is not constrained by the WISE data
alone; this leaves a slight excess in Band~3 whose origin is unclear.


\section{Classical novae observed by WISE}
Like RNe, CN eruptions occur following thermonuclear runaway on the surface
of a WD in a semi-detached binary system. The nature of the WD can determine
the course of the eruption. CN eruptions on CO WD (``CO novae'') tend to result
in the ejection of lower mass than those on ONe WDs, and often result in dust
formation. CN eruptions on ONe WDs (``neon novae'') eject higher mass, and often
result in coronal emission; CNe displaying coronal emission are rarely strong
dust producers. The energetics
of the eruption are characterised by $t_2$ (or $t_3$).
\cite{strope} provide a description of CN LCs, and $t_2$ and $t_3$ for a large
number of objects.

Overviews of CNe are given in \cite{AN,gehrz-pasp,CNII,Woodward-Starrfield2011,basi,stella}.

\subsection{V603 Aql}
This is a well-studied nova with a well-observed stellar and nebular remnant.
It was not detected by \iras\ \citep{HG88}.
There are strong detections in WISE Bands~1 and 2, and good positional coincidence
($0.\!\!''5$) between the WISE source and the nova; we are confident that the WISE
source is associated with the nova remnant. However the WISE data do not enable us
to draw any conclusions about the nature of the IR emission.

\subsection{V1229 Aql}
A poorly-observed nova, with a resolved optical shell \citep{cohen}; it was
not detected by \iras\ \citep{HG88}. The WISE source
is $<1.\!\!''7$ from the nova.
There is a strong detection in WISE Band~1 and a moderate detection in Band~2; the nova
is not detected in Bands~3 and~4. Again no conclusions can be drawn about the nature of
the IR emission.

\subsection{V1370 Aql}
V1370 Aql was a well-observed dusty nova and was one of the first to show the now
familiar ``chemical dichotomy'', in which both C-rich and O-rich dusts condense
\citep{snijders}. In their survey of novae with \iras\ \cite{HG88} reported a detection in \iras\
Band~1, although \cite{callus} give only upper limits.

The WISE source is some $2.\!\!''7$ from the nova, but within the
uncertainties of the WISE positions. As with V1229~Aql no conclusions can be drawn 
about the nature of the IR emission.

\begin{figure*}
\setlength{\unitlength}{1cm}
\begin{center}
\leavevmode
\begin{picture}(5.0,22)
\put(0.0,4.0){\includegraphics{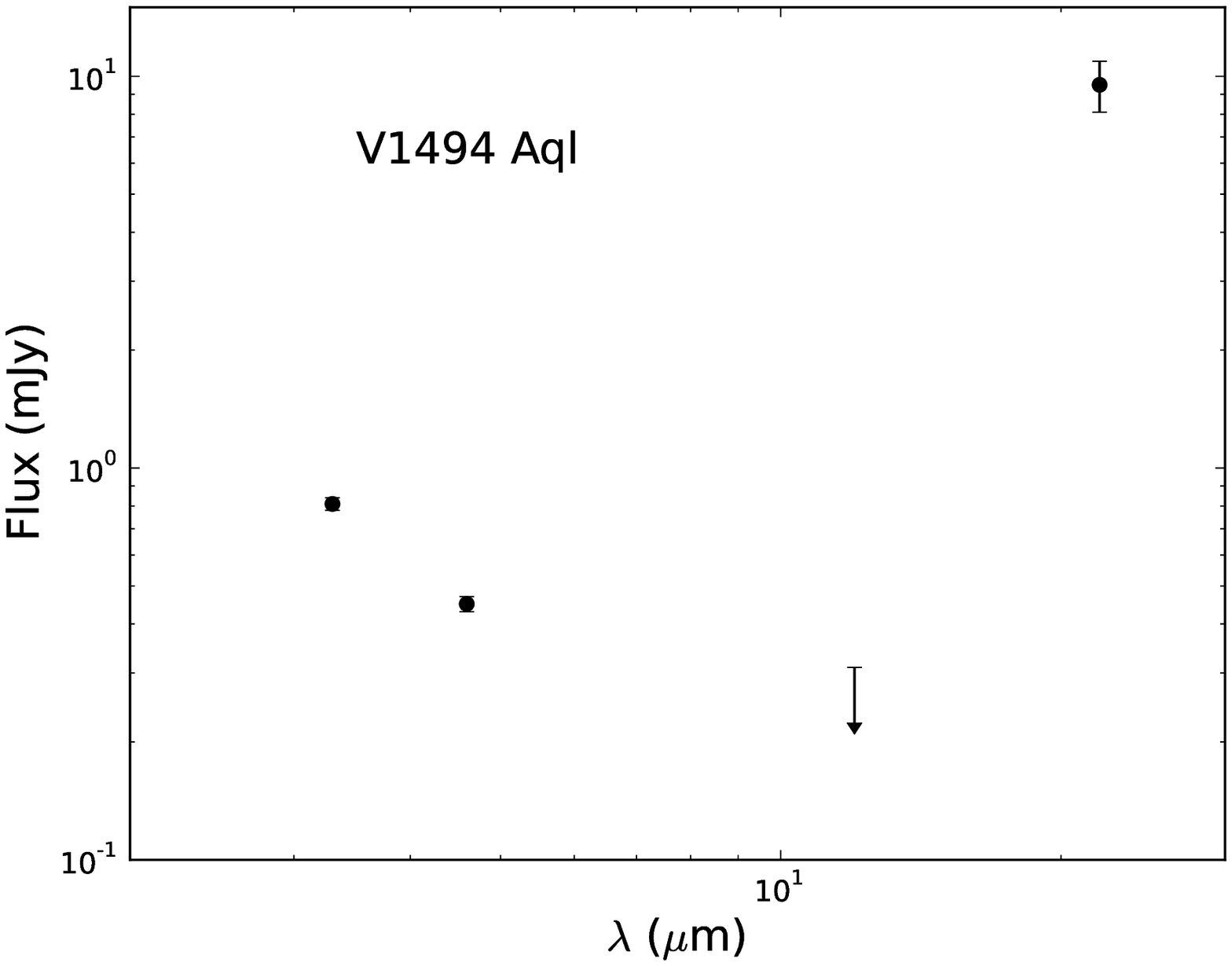}}
\put(0.0,4.0){\includegraphics{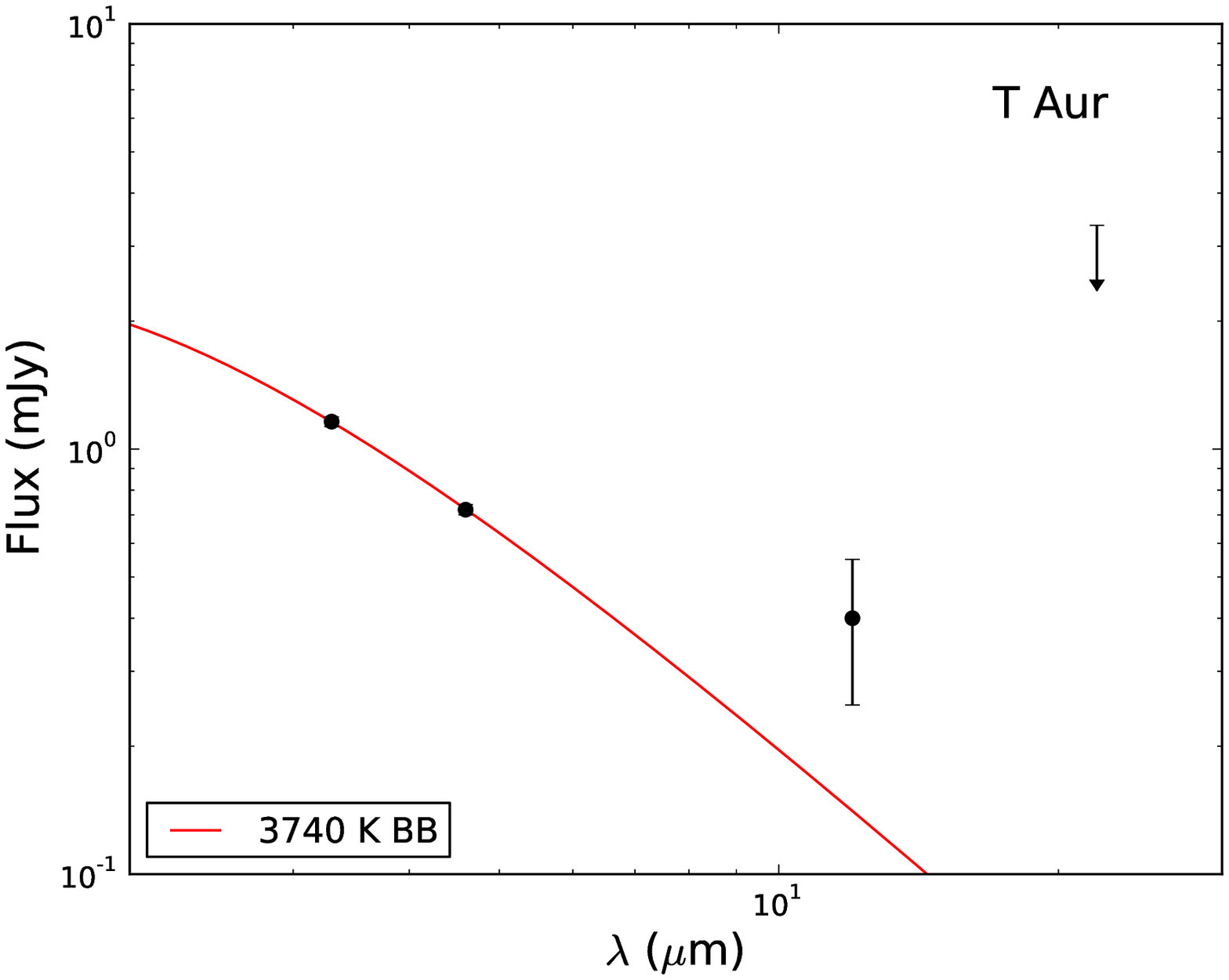}}
\put(0.0,4.0){\includegraphics{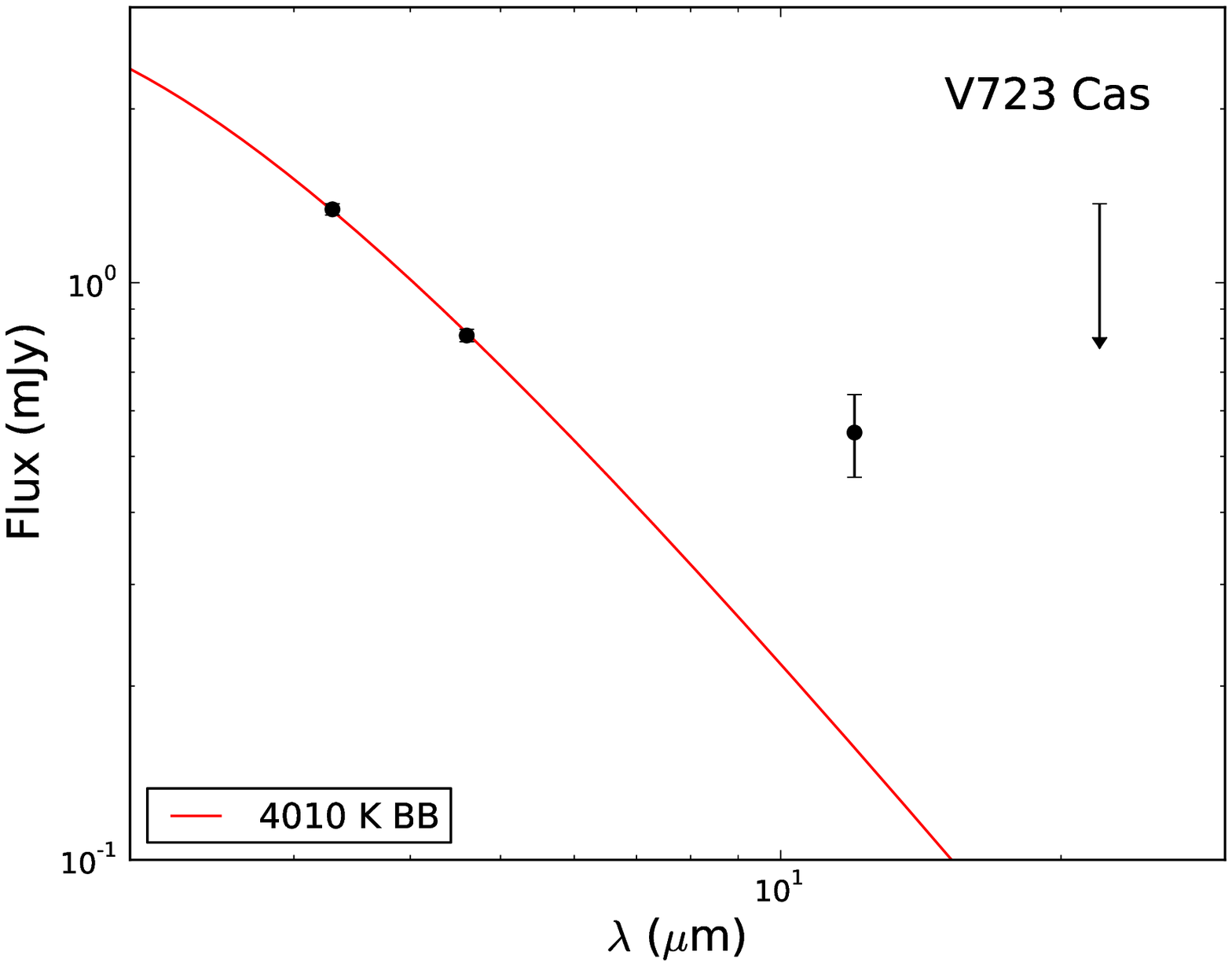}}
\put(0.0,4.0){\includegraphics{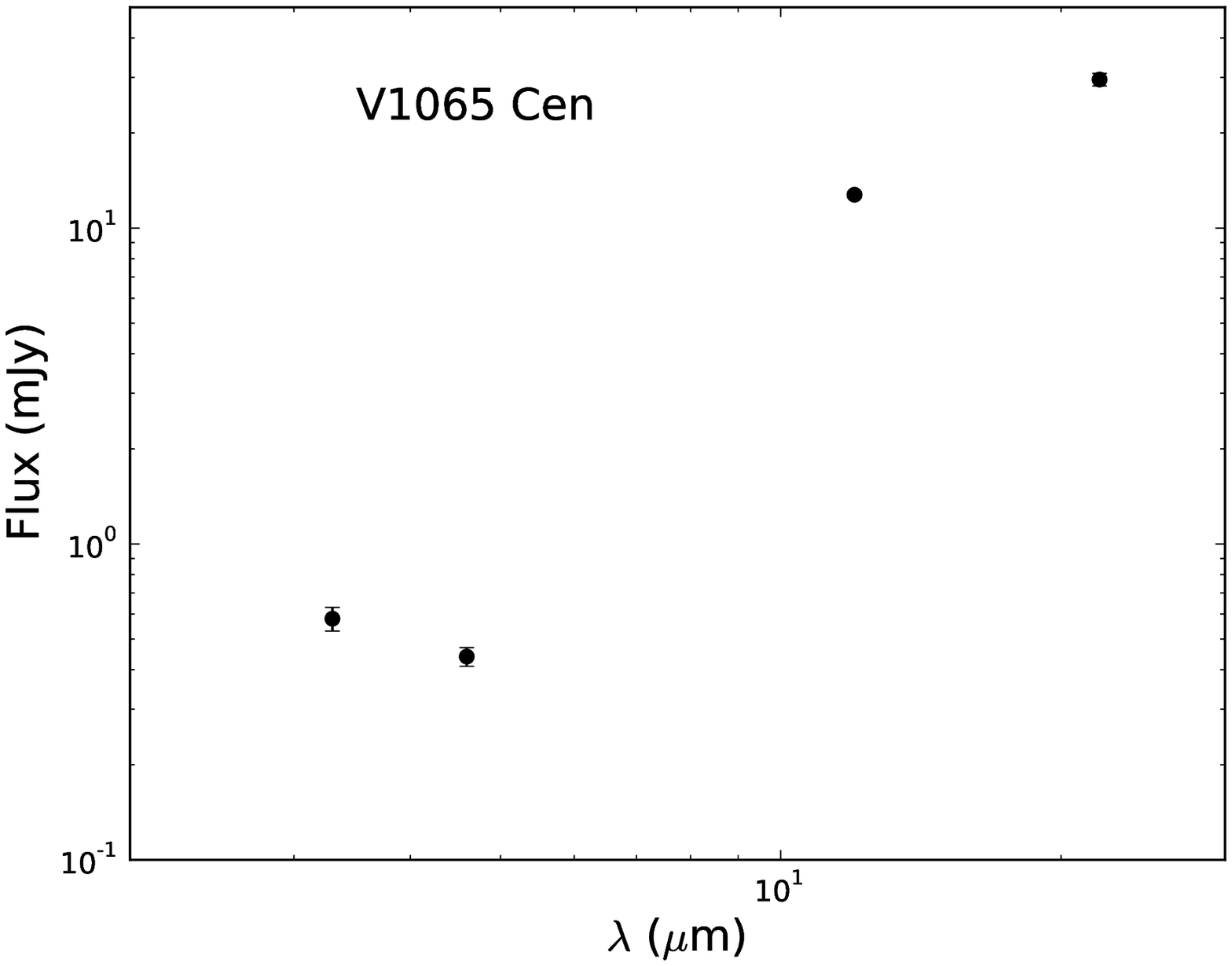}}
\put(0.0,4.0){\includegraphics{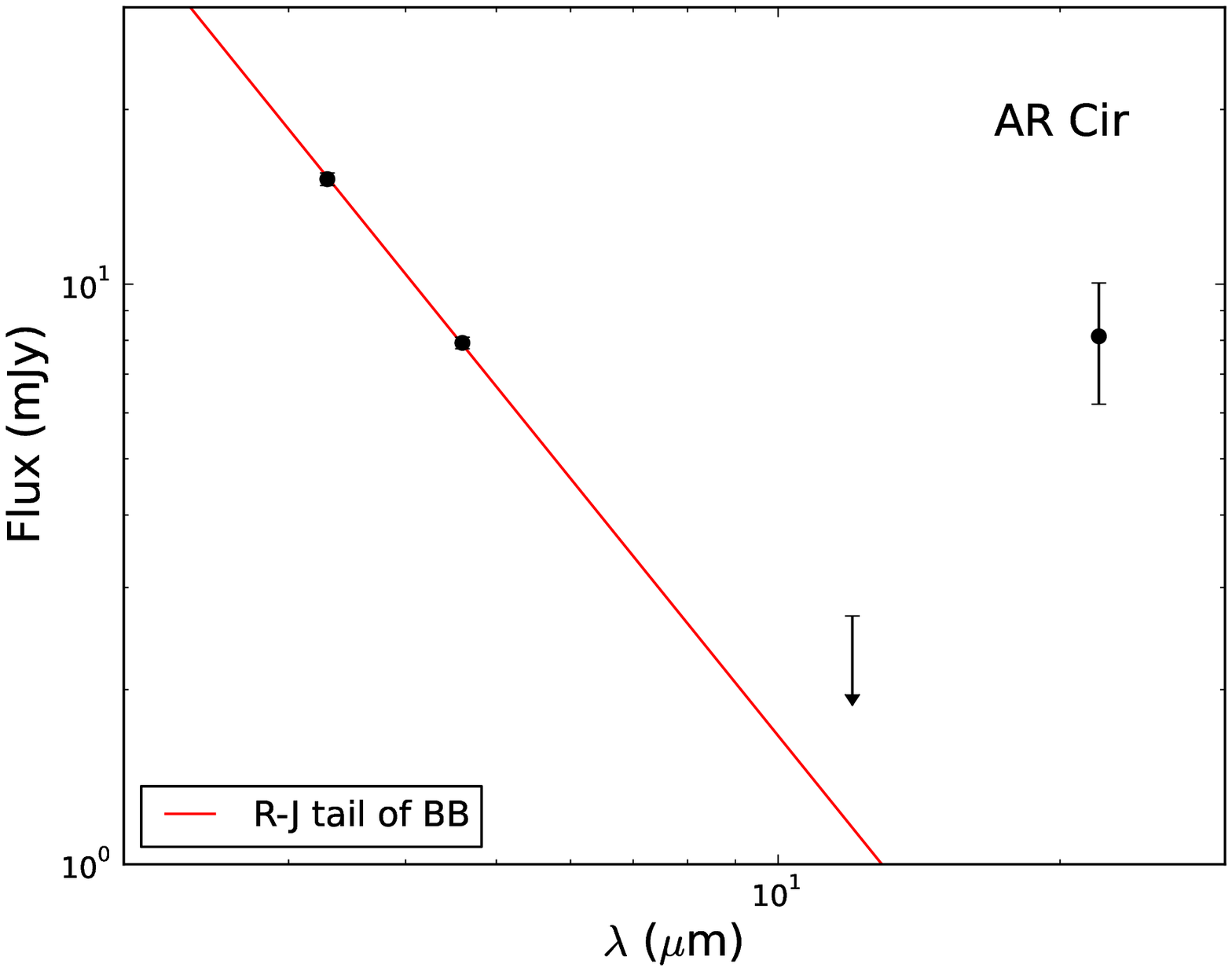}}
\put(0.0,4.0){\includegraphics{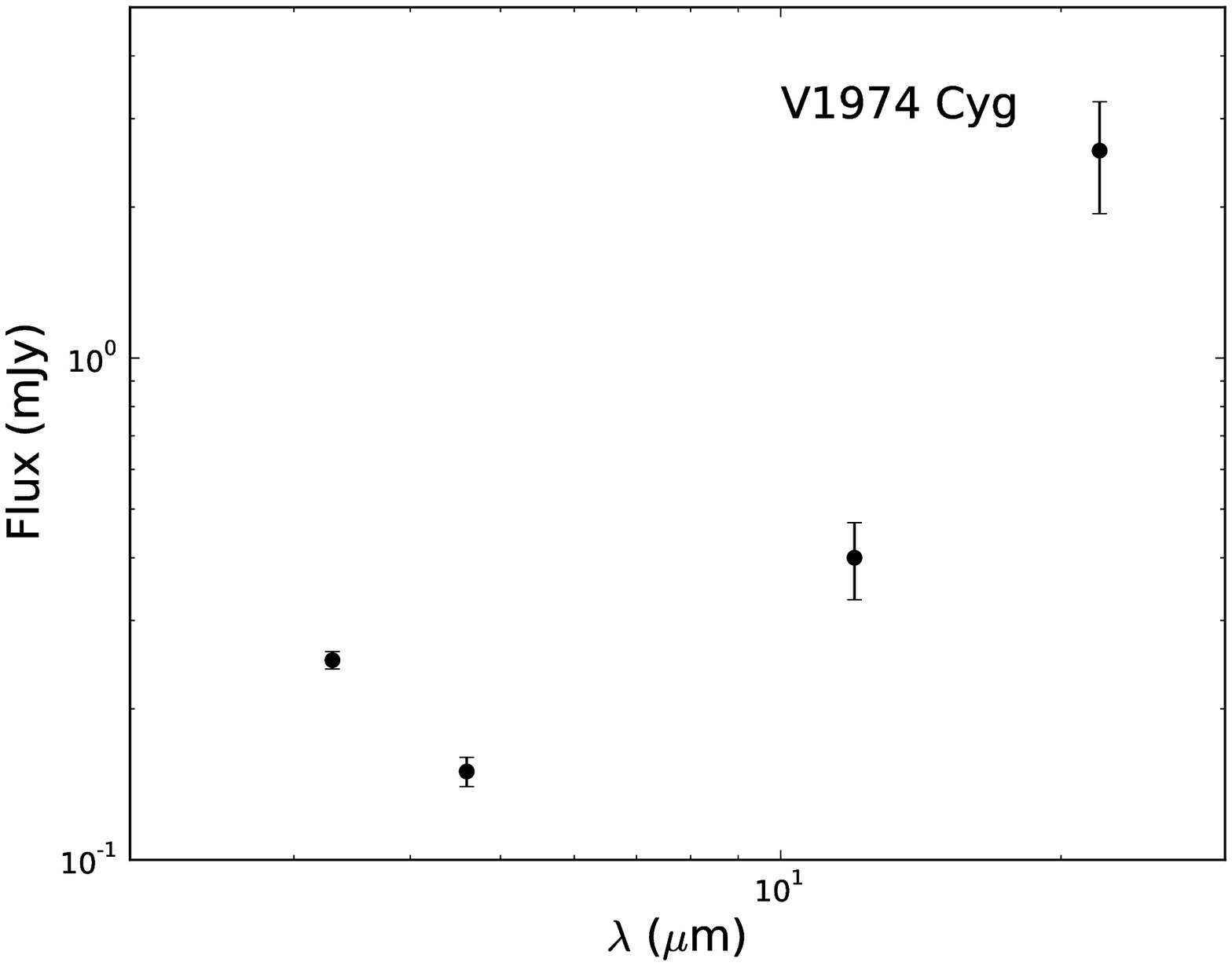}}
\put(0.0,4.0){\includegraphics{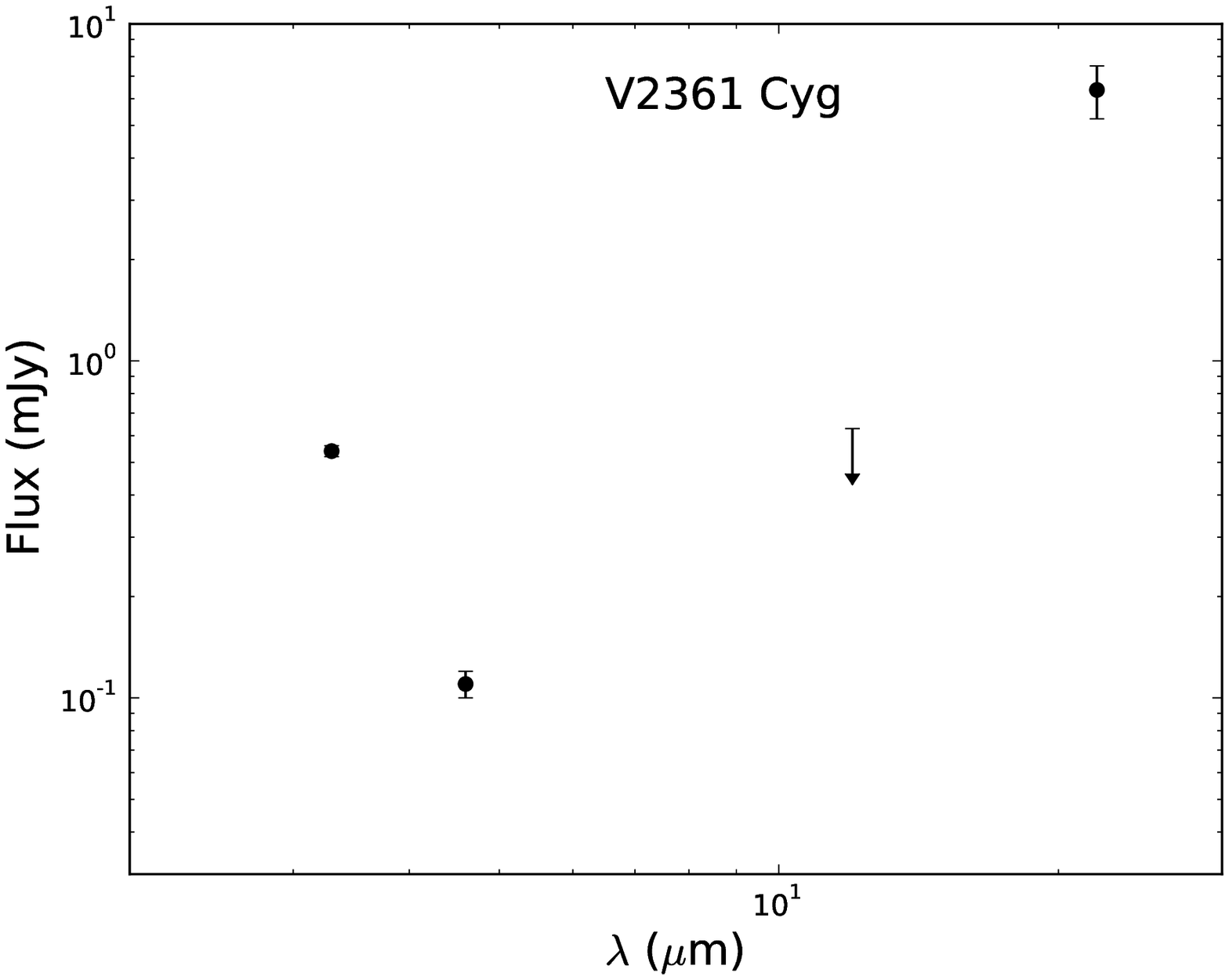}}
\put(0.0,4.0){\includegraphics{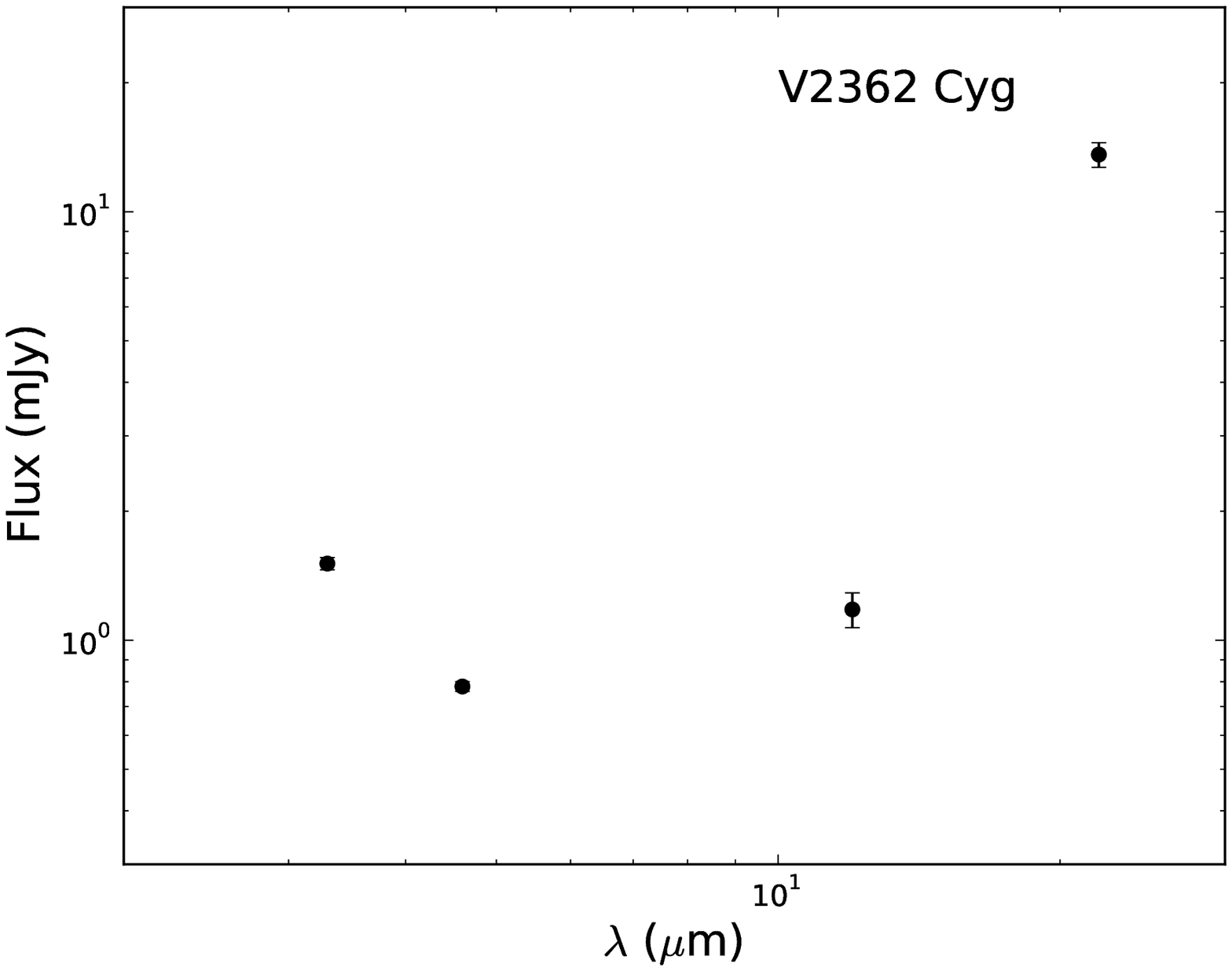}}
\end{picture}
\caption[]{Classical novae in the WISE database; see text for details.
Top left: V1494 Aql.
Top right: T~Aur.
Upper middle left: V723 Cas.
Upper middle right: V1065 Cen.
Lower middle left: AR Cir.
Lower middle right: V1974 Cyg.
Bottom left: V2361 Cyg.
Bottom right: V2362 Cyg.
Errors are smaller than plotted points if not shown. 
\label{CN1}} 

\end{center}
\end{figure*}

\setcounter{figure}{2}

\begin{figure*}
\setlength{\unitlength}{1cm}
\begin{center}
\leavevmode
\begin{picture}(5.0,22.)
\put(0.0,4.0){\includegraphics{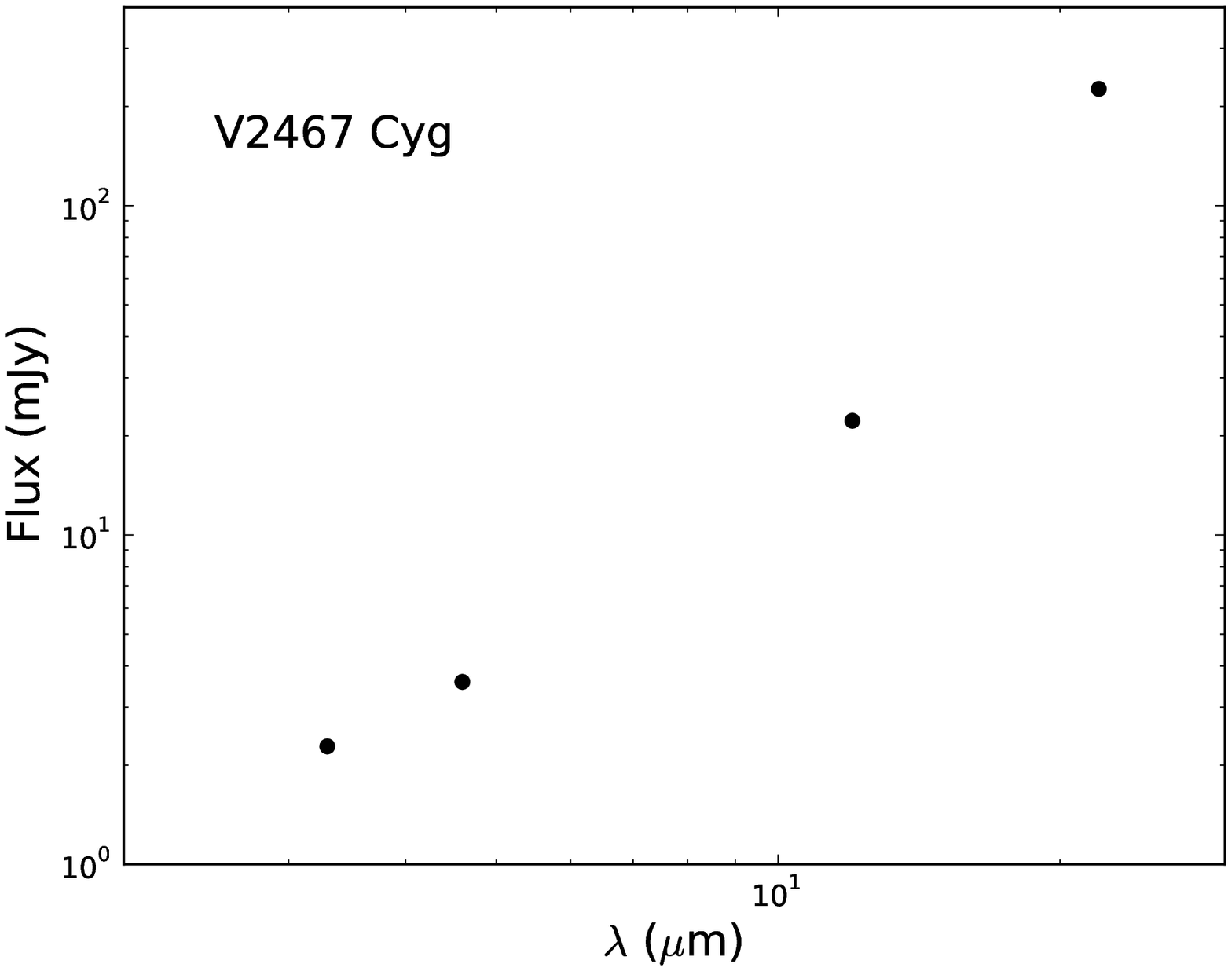}}
\put(0.0,4.0){\includegraphics{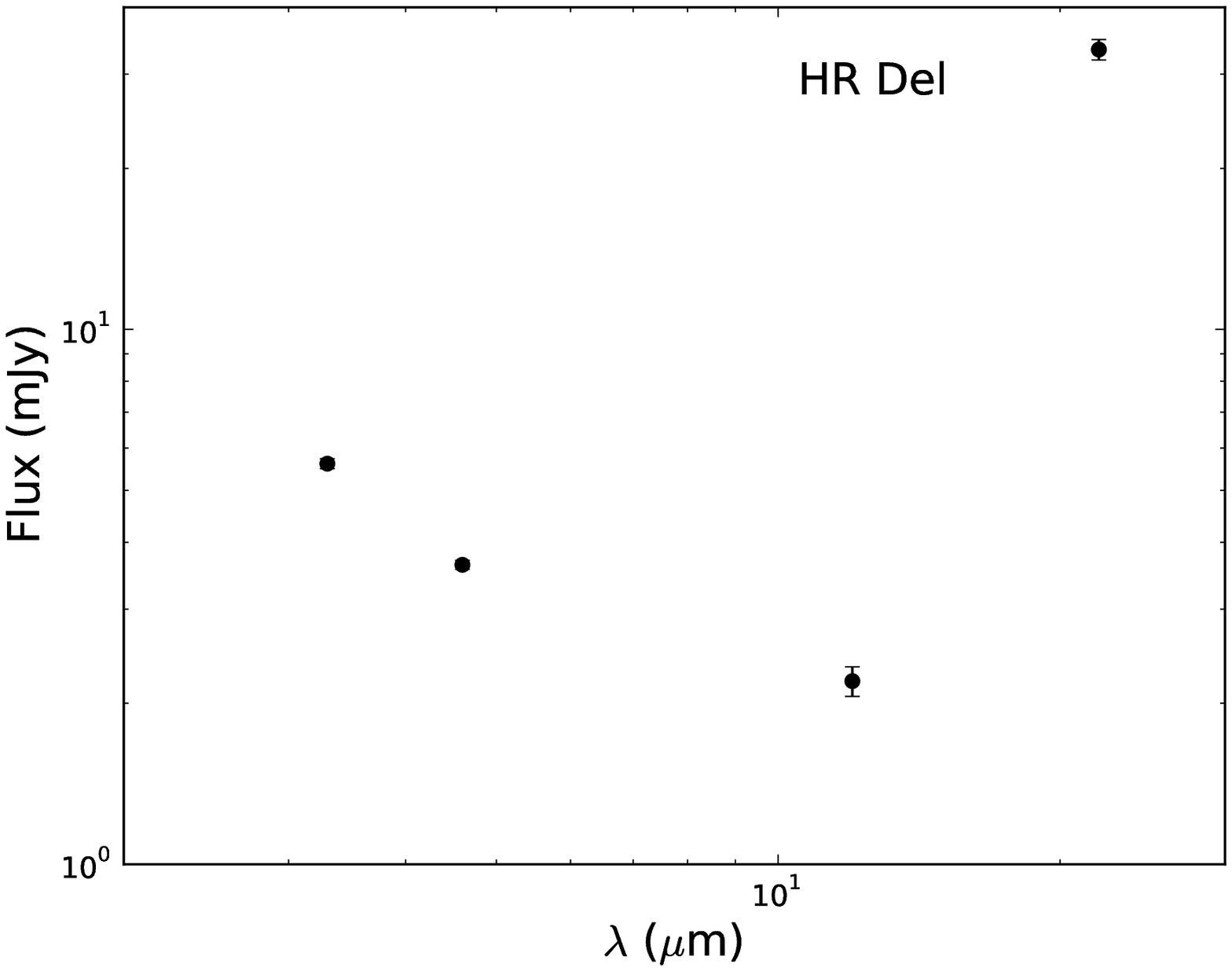}}
\put(0.0,4.0){\includegraphics{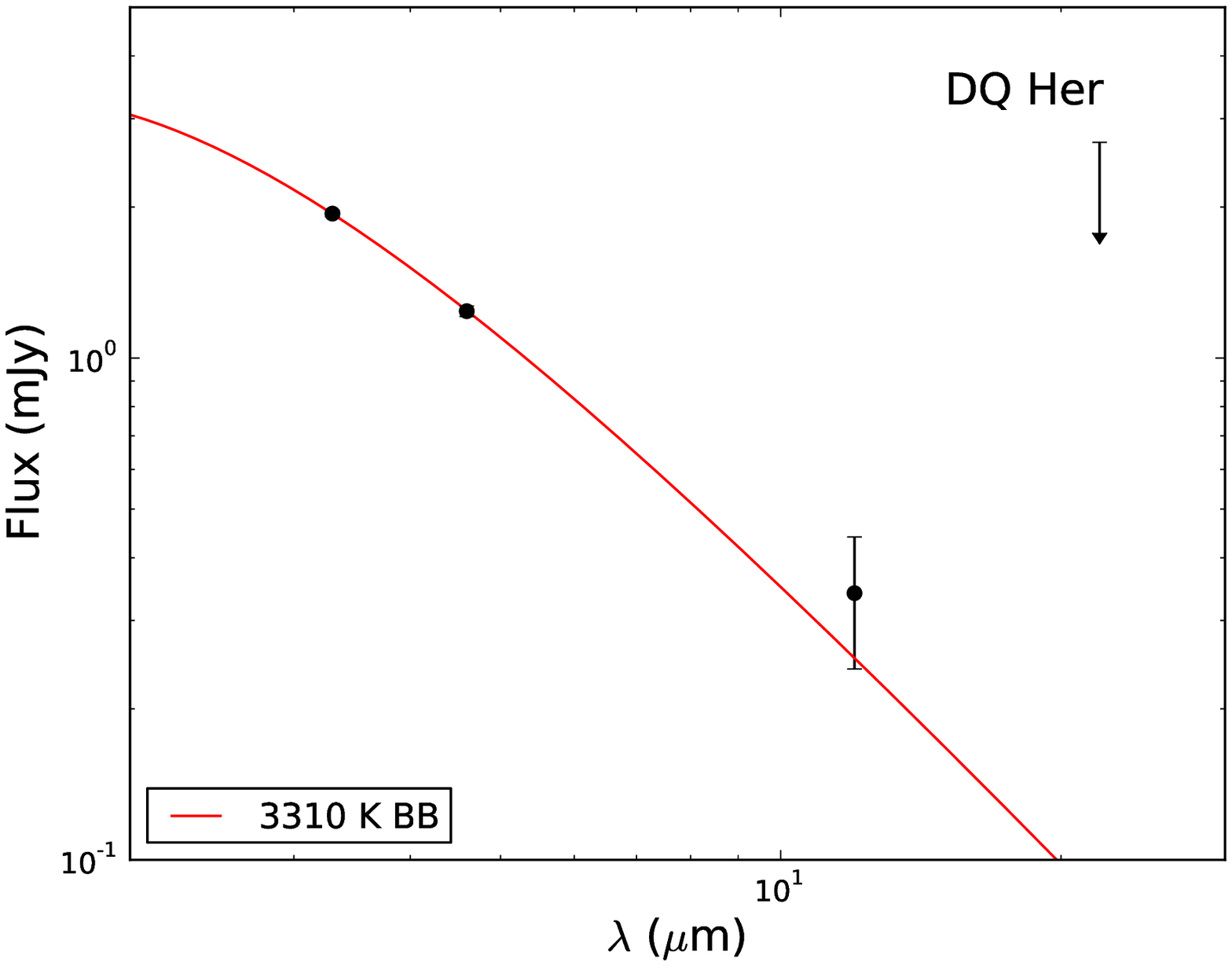}}
\put(0.0,4.0){\includegraphics{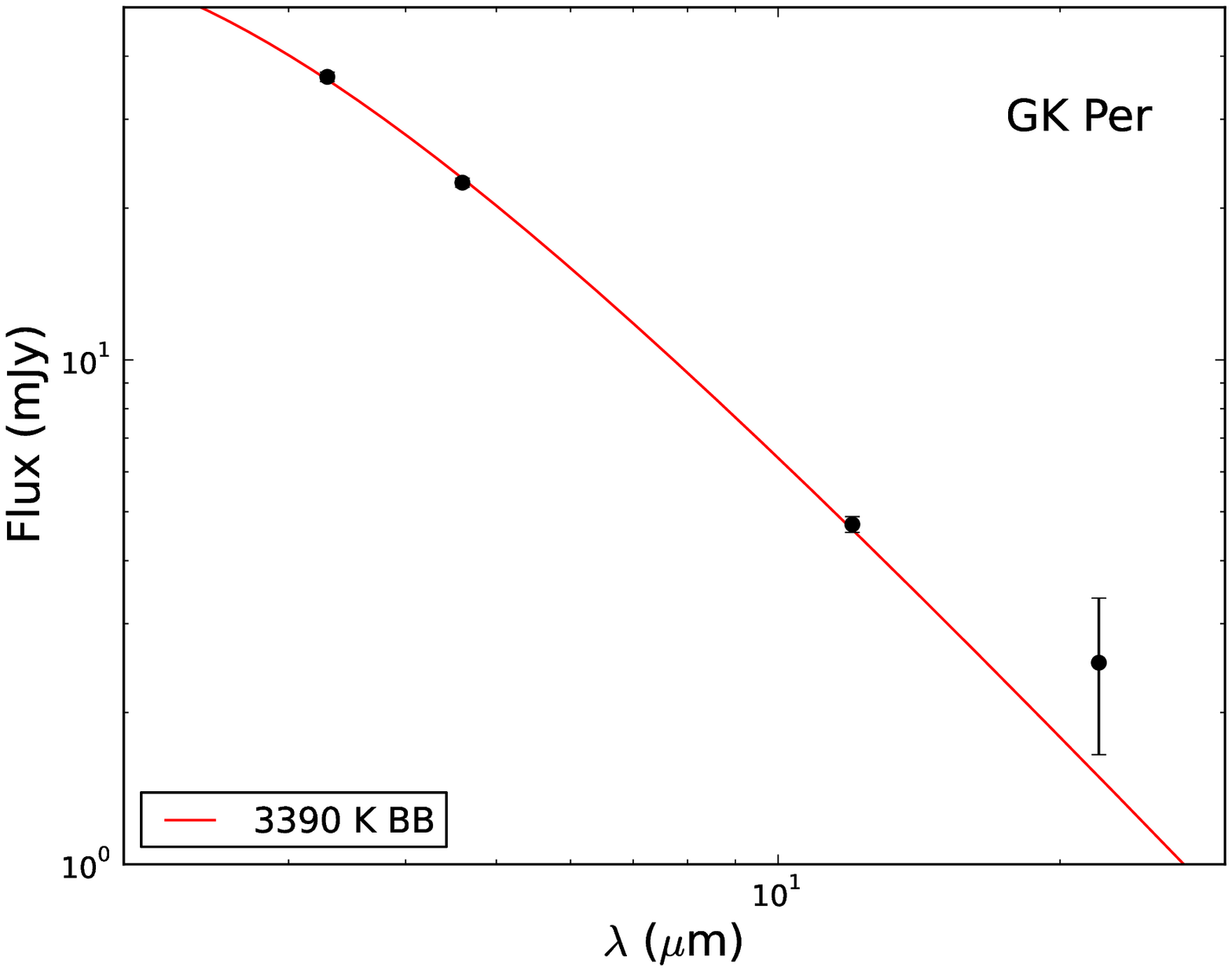}}
\put(0.0,4.0){\includegraphics{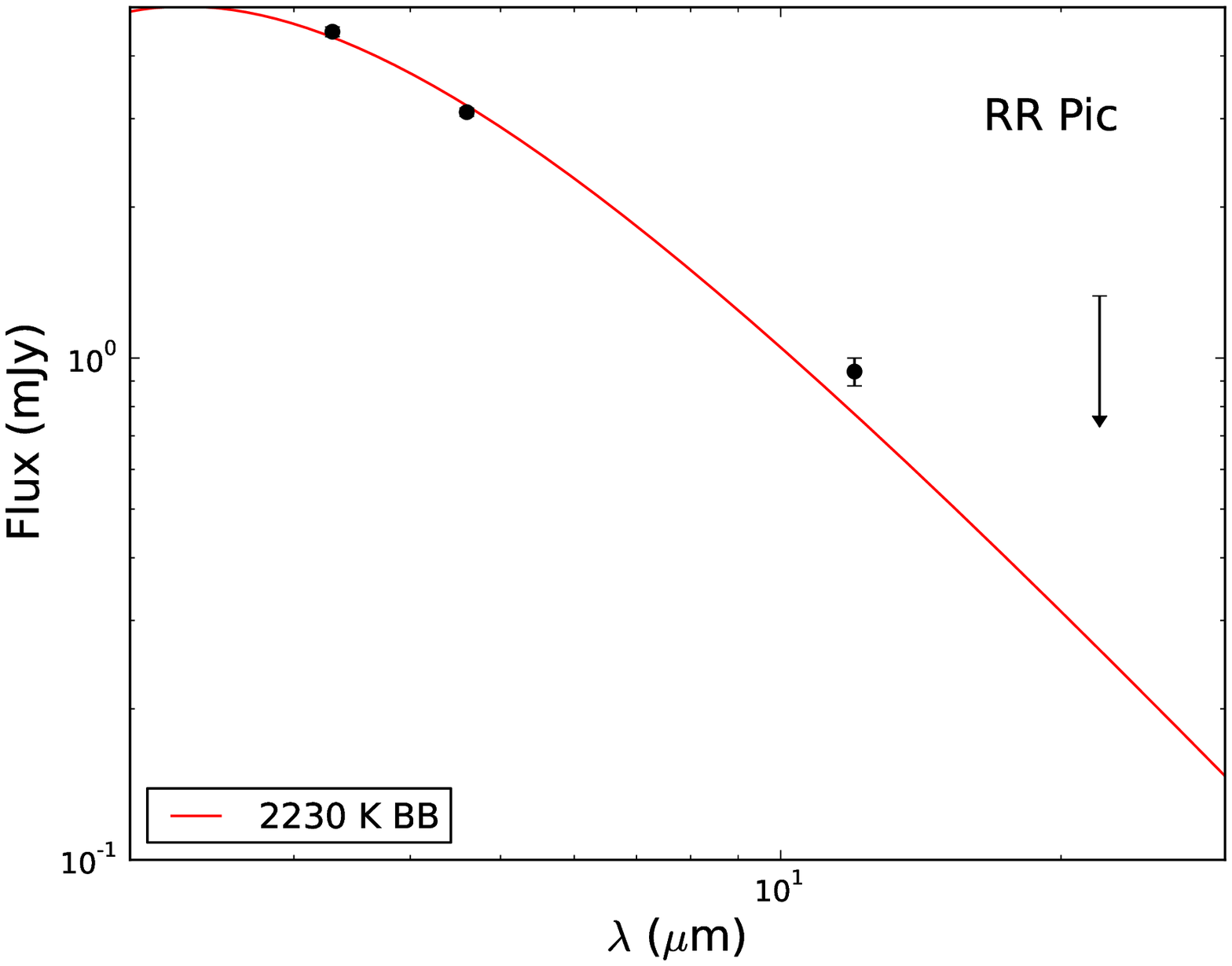}}
\put(0.0,4.0){\includegraphics{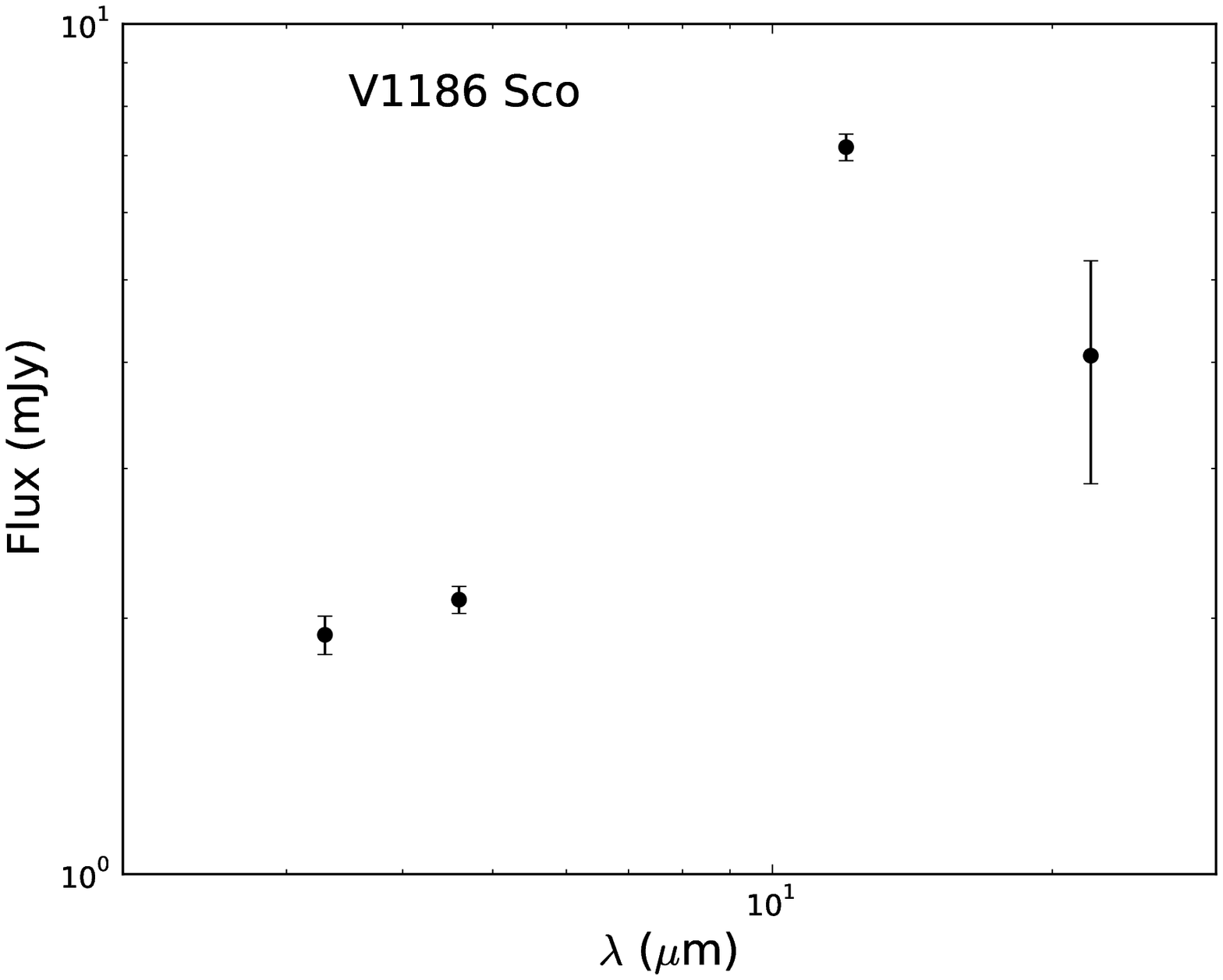}}
\put(0.0,4.0){\includegraphics{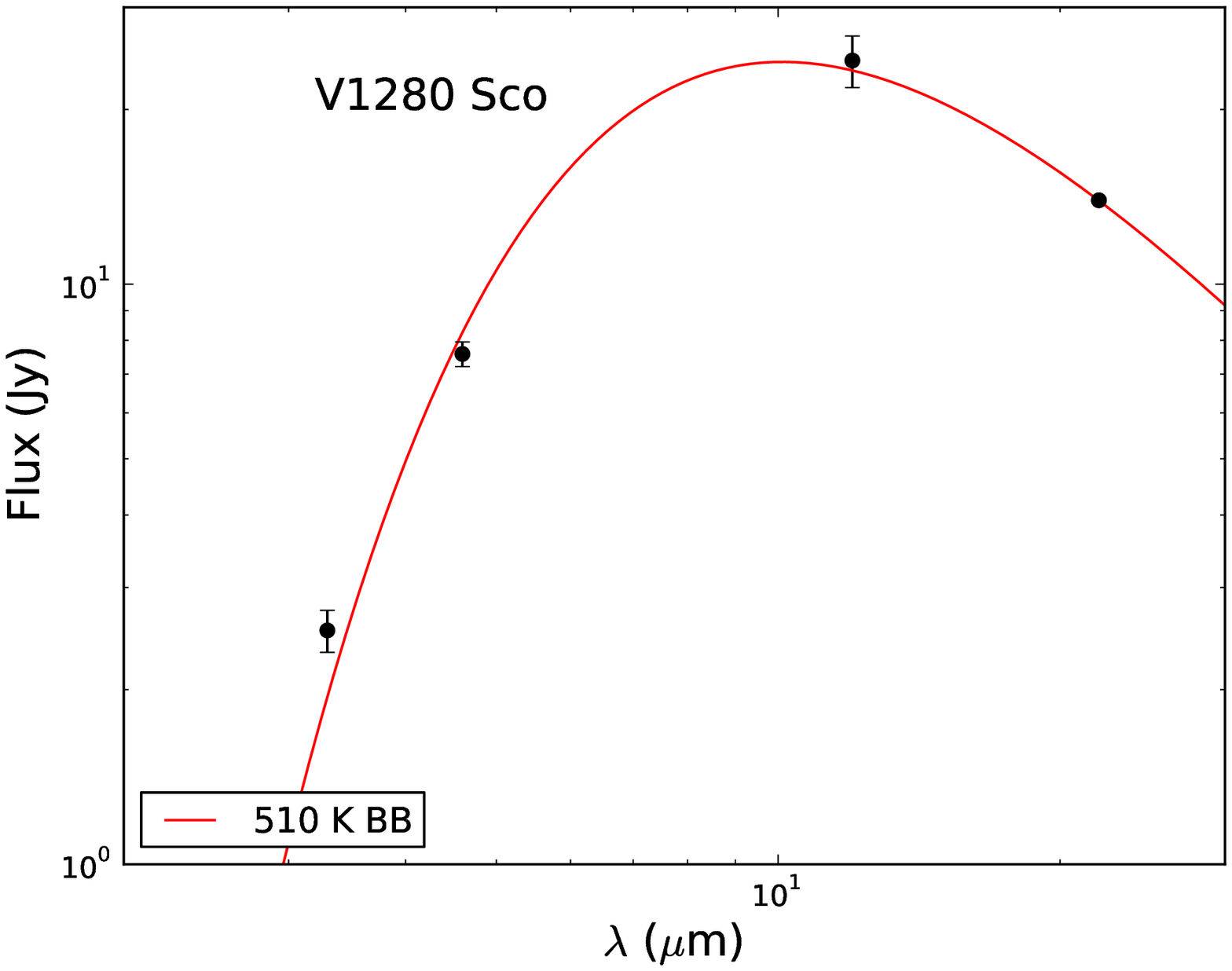}}
\put(0.0,4.0){\includegraphics{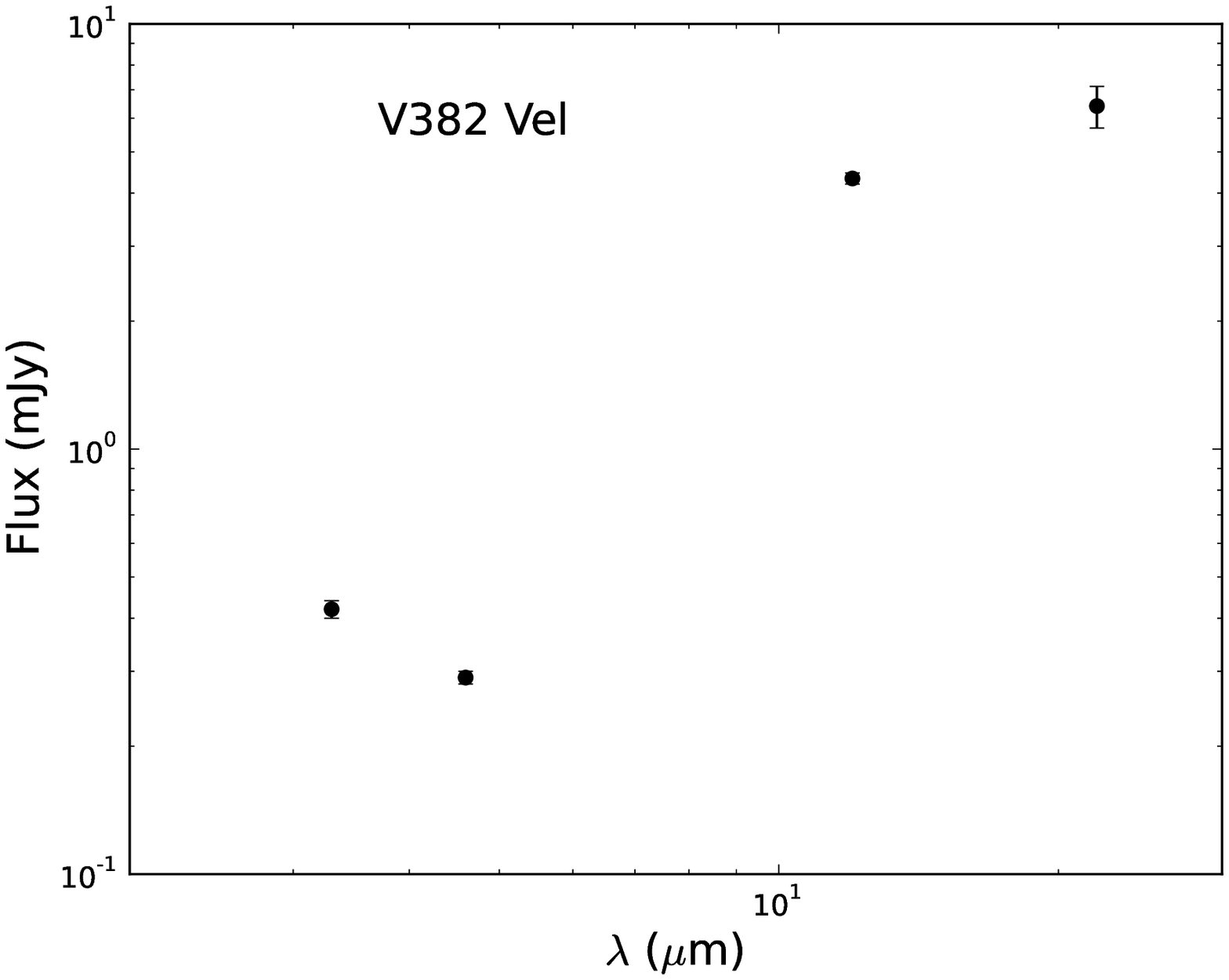}}
\end{picture}
\caption[]{Continued. Classical novae in the WISE database; see text for details.
Top left: V2467 Cyg.
Top right: HR Del.
Upper middle left: DQ Her.
Upper middle right: GK Per.
Lower middle left: RR Pic.
Lower middle right: V1186 Sco.
Lower left: V1280 Sco.
Lower right: V382 Vel.
Errors are smaller than plotted points if not shown.
\label{CN2}} 
\end{center}
\end{figure*}

\subsection{V1494 Aql}
This was a very well-observed nova in which the binary displays eclipses. Optical and radio
observations \citep{eyres} showed the material ejected in the 1999 eruption was highly
clumpy and non-spherical, while X-ray observations \citep{drake} revealed oscillations 
and a transient ``burst''.

V1494~Aql was observed using the \spitzer\ IRS by \cite{helton}, who complemented the
IR data with optical spectroscopy. The flux in the \fion{O}{iv}~25.9\mic\ fine structure
line was $18.7\times10^{-16}$~W~m$^{-2}$ in 2007, with a peak flux density of $\sim6$~Jy
over the period 2004 April -- 2007 October \citep{helton}. Oxygen was overabundant relative
to solar by a factor of $\gtsimeq15$, depending on the assumed electron density.

The WISE SED is included in Fig.~\ref{CN1}, in which the largest flux is in Band~4. Given the
strength of \fion{O}{iv}~25.9\mic\ in 2007 it seems probable that the flux in the WISE
filters likely has an emission line origin.

\begin{table*}
\caption{Pre-2010 novae not detected by WISE. \label{non-det1}}  
\begin{tabular}{lclc}  \hline
Nova & Outburst year & Nova & Outburst year \\ \hline
LS And    & 1971   & V838 Her  & 1991  \\     
OS And    & 1986   & CP Lac    & 1935  \\
CI Aql    & 1917   & BT Mon    & 1939  \\ 
V1301 Aql & 1975   & CP Pup    & 1942  \\ 
V1419 Aql & 1993   & V597 Pup  & 2007  \\ 
V842 Cen  & 1986   & V4077 Sgr & 1982  \\
IV Cep    & 1971   & V4021 Sgr & 1977  \\
BY Cir    & 1995   & V723 Sco  & 1952  \\
Q Cyg     & 1876   & V1187 Sco & 2004 \\
V1668 Cyg & 1978   & FH Ser    & 1970  \\
V2274 Cyg & 2001   & LW Ser    & 1978  \\
V2361 Cyg & 2005   & NQ Vul    & 1976  \\
V2491 Cyg & 2008   & PW Vul    & 1984  \\
DN Gem    & 1912   & QU Vul    & 1984  \\ 
V446 Her  & 1960   & QV Vul    & 1987  \\
V827 Her  & 1987   &           &       \\
\hline\hline 
\end{tabular}
\end{table*}

\subsection{T Aur}
T Aur is another nova system in which the central binary is eclipsing; there was no
\iras\ source at the position of the nova \citep{HG88}. The WISE
source is within $\sim0.\!\!''2$ of the nova; the WISE data are shown in Fig.~\ref{CN1}.
The data in Bands 1 and 2
are consistent with a photospheric blackbody at 3\,740~K, suggesting that the
WISE Band~1, 2 data are seeing the main sequence secondary.
However this leaves an excess at 12\mic\ whose origin is not clear.

\subsection{QZ Aur}
Another poorly-observed nova with a resolved optical shell; the WISE source
is within $0.\!\!''2$ of the nova position.
We have detections in WISE Bands~1 and~2 only, and no conclusions can be drawn
about the nature of the IR emission.


\subsection{T Boo}
T~Boo is the oldest ``old nova'' for which we have searched the WISE database. The WISE source
is some $1.\!\!''7$ from the position of the nova, well within the WISE positional uncertainty.
There are strong detections in Bands~1 and~2, with a marginal detection in Band~3.
Without complementary data it is not possibly to determine the origin of the IR emission.

\subsection{V705 Cas}

\begin{figure*}
\setlength{\unitlength}{1cm}
\begin{center}
\leavevmode
\begin{picture}(5.0,9.5)
\put(0.0,4.0){\includegraphics{v705cas.eps}}
\put(0.0,4.0){\includegraphics{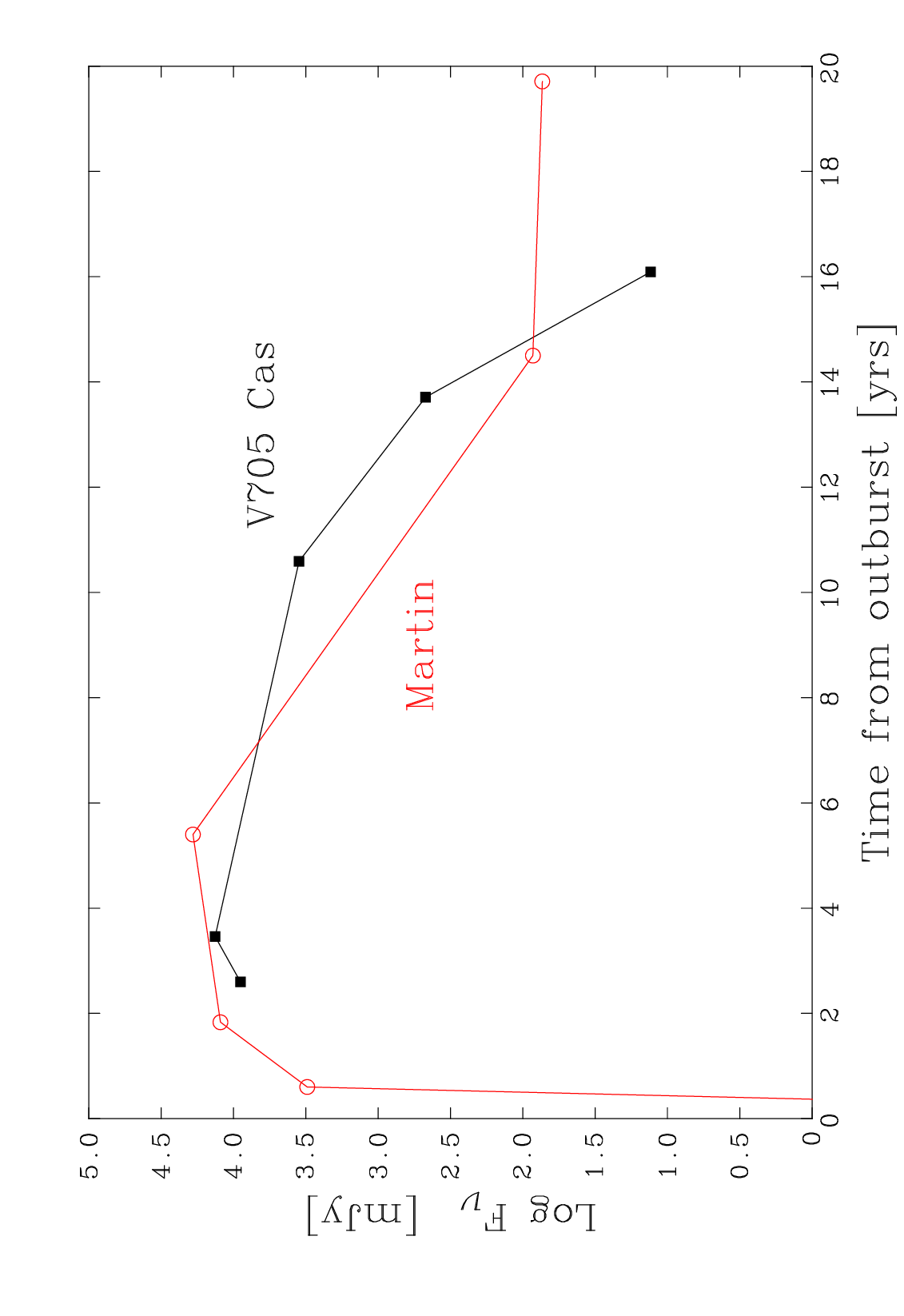}}
\put(0.0,4.0){\includegraphics{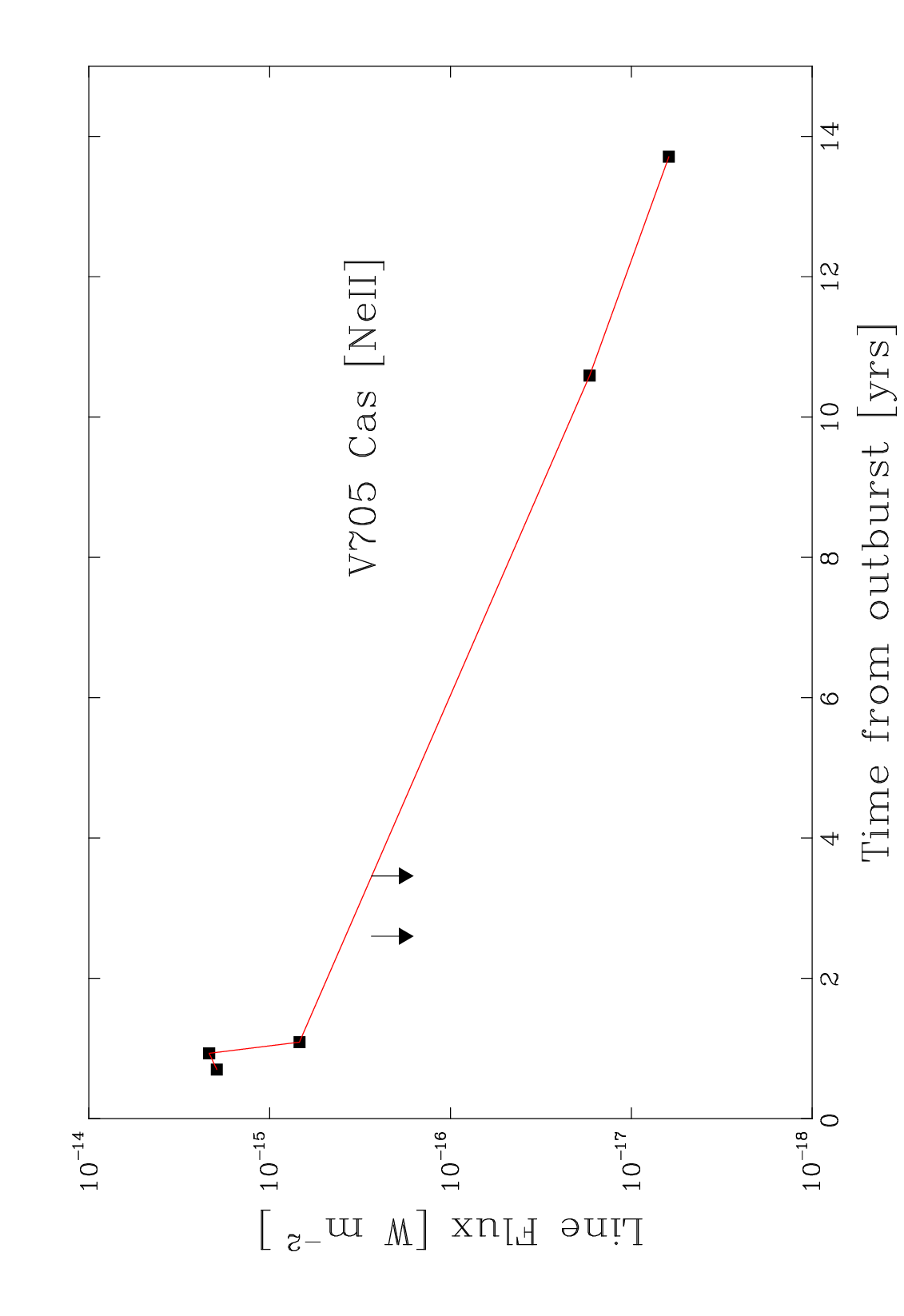}}
\end{picture}
\caption[]{Top: WISE SED of V705Cas.
Bottom left: Black points and line: peak flux (in Jy) of the \fion{O}{iv}~25.9\mic\
fine-structure line in V705Cas, based on \iso, \spitzer\ and WISE data; red points and
line: predicted variation of the \fion{O}{iv} flux, calculated for DQ~Her by \cite{martin}.
Bottom right: observed variation of the \fion{Ne}{ii}~12.8\mic\ fine structure line,
based on ground-based \citep{evans-cas2},
\iso\ \citep{salama} and \spitzer\ IRS (Helton et al., in prepration) data.
See text for details.
\label{v705cas}} 
\end{center}
\end{figure*}

This dusty nova was discovered on 1993 December 14 and was well-observed in the IR.
It exhibited a deep minimum in its visual LC, the characteristic signature
of prolific dust formation \citep[][Helton et al., in
preparation]{mason,evans-cas1,evans-cas2,evans-cas4,salama};
maximum dust emission occurred around day~105 \citep{mason}. Observations using 
the Short Wavelength Spectrometer (SWS) on the \iso\ detected no dust, although
\spitzer\ IRS observations (Helton et al., in preparation) show a weak dust
continuum longward of $\sim15$\mic. We note however that the ISO observations were
executed some time (days 950, 1265 and 1455) after maximum dust emission.

The \fion{O}{iv}~25.9\mic\ fine structure line
was clearly detected with \iso\ in observations made 950~days and 1265~days
after eruption, with peak flux 8.94~Jy and 13.4~Jy respectively, and integrated
flux $5.2\times10^{-15}$~W~m$^{-2}$ and $5.8\times10^{-15}$~W~m$^{-2}$
respectively \citep{salama}. \citeauthor{salama} concluded that O was over-abundant in V705~Cas
by at least 25 relative to solar, while the neon abundance was $0.5\times$ solar
at most. The \fion{O}{iv} line was also strongly detected by the \spitzer\ IRS
(Helton et al., in preparation).

The WISE source is within $0.\!\!''2$ of the nova and the SED is included in Fig.~\ref{v705cas}.
In WISE Bands 1--3 the flux is
consistent with a weak continuum but there is a clear excess in Band~4, with a
peak flux density of 13.08~mJy. This is very likely due to emission in the
fine-structure \fion{O}{iv}~25.9\mic\ line, which was prominent in the first
$\sim3$~years after outburst.

On the assumption that the Band~4 flux is indeed due
to \fion{O}{iv}, the decline in the peak flux in the \fion{O}{iv}
line in V705~Cas over the period 1996 (\iso) -- 2007 (WISE) is shown in
Fig.\ref{v705cas}. In this figure we take the peak flux in the line as a proxy
for the integrated flux, and we make the implicit assumption that there is no
change in line width with time.
In a comprehensive study of the evolution of the line emission
of nova DQ~Her (1934), \cite{martin} undertook a detailed study of the
long-term ($\sim50$~years) evolution of nebular, coronal and fine-structure lines;
the study included a number of the mid- and far-IR lines
now routinely seen in the spectra of mature novae. The predicted evolution
of the \fion{O}{iv} flux is also shown in Fig.~\ref{v705cas}. With the caveat that
\citeauthor{martin}'s analysis was for the specific case of DQ~Her, its stellar
remnant and ejecta abundances, the predicted evolution of the \fion{O}{iv} fine
structure line is in surprisingly good (qualitative) agreement with the evolution
of the \fion{O}{iv} line in V705~Cas.

While the \fion{Ne}{ii}~12.8\mic\ fine structure line was detected in
ground-based observations \citep{evans-cas2} and by the \spitzer\ IRS (Helton et al.,
in preparation), it was not detected by \iso. We also show
in Fig.~\ref{v705cas} the observed evolution of the \fion{Ne}{ii}~12.8\mic\
fine structure line in V705~Cas over a $\sim15$~year period. It is evident that the
\fion{Ne}{ii} line in V705~Cas peaked much earlier, and decayed more quickly,
than the \fion{O}{iv} line.

\vspace{-3mm}

\subsection{V723 Cas}
This nova had an erratic LC, and displayed strong coronal
line emission in the IR \citep{evans-v723}; the distance is 4~kpc \citep{evans-v723,lyke}.
\citeauthor{evans-v723} also concluded that
the secondary must be evolved for mass transfer to occur. \cite{lyke} found that
high spatial resolution images
showed the \fion{Si}{vi} and \fion{Ca}{viii} coronal lines form an
equatorial ring, while \fion{Al}{ix} was in a prolate spheroid; all of these lines were
detected from ground-based IR observations by \cite{evans-v723}.

The WISE source is within $0.\!\!''2$ of the position of the nova; the WISE
SED is included in Fig.~\ref{CN1}. The fluxes in Bands 1 and 2 fit well to a
blackbody with temperature 4\,010~K and radius 5.5\Rsun, reasonably consistent with a K2IV
(effective temperature 4\,600~K) star at 4~kpc.
We therefore tentatively conclude that the WISE data are consistent with photospheric emission
and, given the absence of dust immediately after outburst \citep{evans-v723}, the Band~3
flux is almost certainly due to line emission.


\subsection{V1065 Cen} V1065~Cen was another dust former and was observed
with the \spitzer\ IRS by \cite{v1065cen}. Analysis of the emission line
spectrum lead to abundances relative to solar of He/H = $1.6\pm0.3$, N/H =
$144\pm34$, O/H = $58\pm18$, and Ne/H = $316\pm58$; the extent of its neon
over-abundance places V1065~Cen in the category of ``neon novae''. The
condensed dust consisted primarily of silicates, and the dust mass is estimated
by \citeauthor{v1065cen} to be in the range $(0.2-3.7)\times 10^{-7}$\Msun.
A sequence of \spitzer\ observations showed that the peak in the dust emission
shifted from $\sim10$\mic\ to $\sim20$\mic, with nebular and fine-structure lines
superimposed, over a $\sim700$~day period \citep{v1065cen}; in particular, the
\fion{Ne}{ii}~12.8\mic, and the \fion{Ne}{v}~24.3\mic\ and \fion{O}{iv}~25.9\mic\
lines were prominent, superimposed
on the broad 9.7\mic\ and 18\mic\ silicate features resectively.

The WISE SED is included in Fig.~\ref{CN1}. In view of the known mid- and far-IR properties
of this nova it is likely that the WISE emission consists of emission lines superimposed on
the still-cooling dust continuum.

\subsection{AR Cir}
\cite{harrison} has suggested on the basis of its 1906 LC that the nova status
of the object is doubtful. Its near-IR colours are more akin to a giant (M0III)
donor star and coupled with its LC and low outburst amplitude,
\citeauthor{harrison} concluded that AR~Cir is more likley a symbiotic star
\citep[see also][]{pagnotta}.

The WISE source is within $0.\!\!''3$ of the target and the WISE SED is shown in
Fig.~\ref{CN1}. All we can say is that the WISE Band~1 and 2 data are consistent with the
Rayleigh-Jeans tail of a blackbody.
The WISE data are insufficient to allow us to draw any
conclusion about the nature of the WISE Band~4 emission.

\subsection{DZ Cru}

\begin{figure*}
\setlength{\unitlength}{1cm}
\begin{center}
\leavevmode
\begin{picture}(5.0,5)
\put(0.0,4.0){\includegraphics{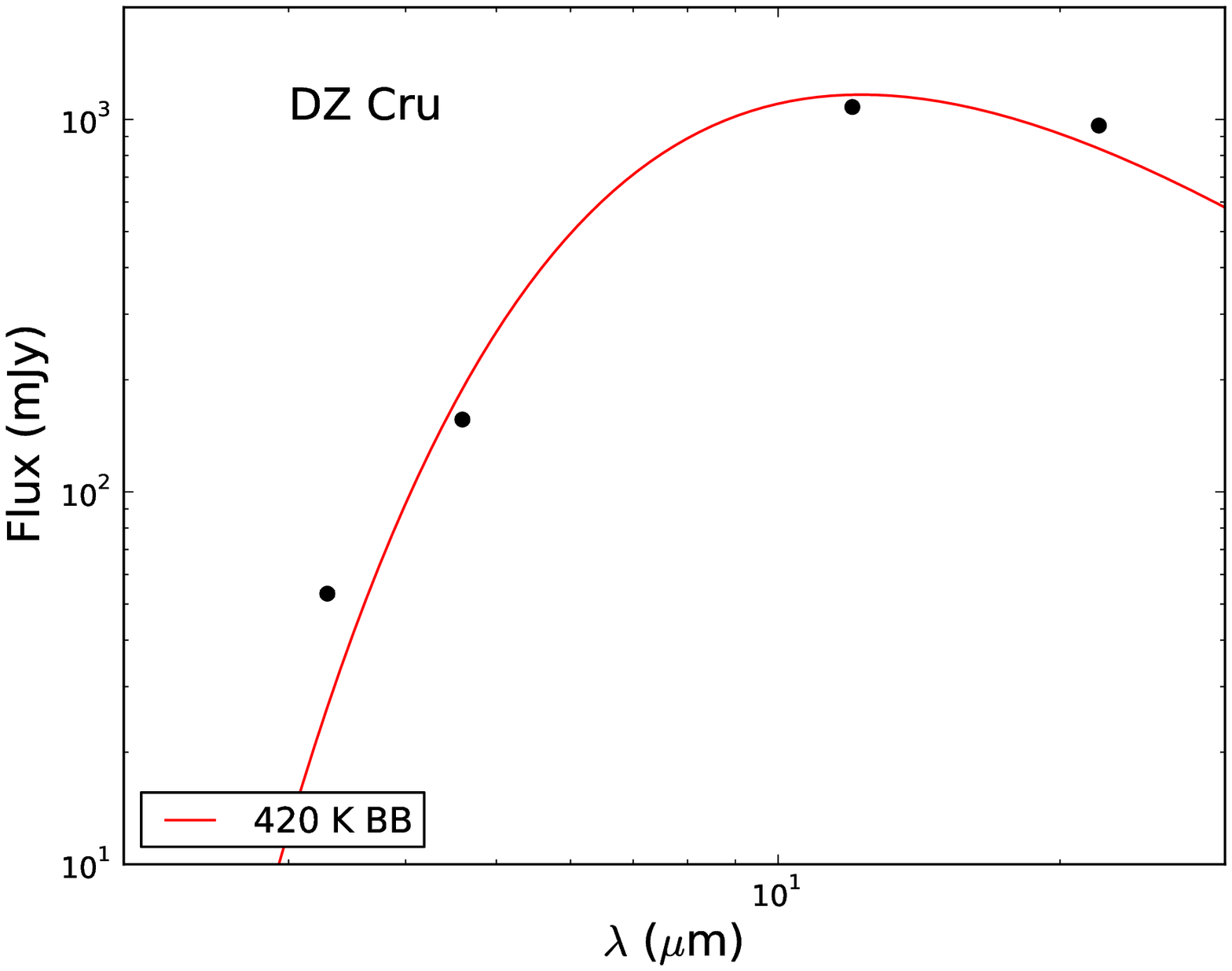}}
\put(0.0,4.0){\includegraphics{dzcru_a.eps}}
\end{picture}
\caption[]{Left: SED of DZ~Cru from WISE. Solid curve is a 420~K blackbody.
Right: Evolution of the dust shell of DZ~Cru, comparing \spitzer\ data
from \cite{evans-dz} (solid spectra) with the WISE data (points). 
The broken curves are 470~K and 420~K blackbodies through the \spitzer\ and WISE data.
Errors are smaller than plotted points.
See text for discussion.
\label{dzcru}} 
\end{center}
\end{figure*}

DZ~Cru was a dusty nova with prominent ``Aromatic Infrared'' (AIR) features
typical of novae. Ground-based IR observations of DZ Cru were carried out by
\cite{rushton}, and \spitzer\ observations by \cite{evans-dz}.

The WISE SED is shown in Fig.~\ref{dzcru}; data from the \spitzer\ campaign
\citep{evans-dz} are shown in a separate frame. The WISE data clearly show the
continued cooling of the dust shell. 
The \spitzer\ data approximately correspond to a 470~K blackbody, while a reasonable
representation of the WISE data is
obtained for a blackbody at $T_{\rm d} = 420$~K (see Fig.~\ref{dzcru}).
For simple outflow the dust temperature should decline with time $t$ as
$T_{\rm d} \propto t^{-2/(\beta+4)}$, where the emissivity of the dust is
$\propto \lambda^{-\beta}$. With $T_{\rm d}\simeq
470$~K when DZ~Cru was observed by \spitzer\ in 2007 September (day 874),
uniform outflow would lead us to expect $T_{\rm d}\simeq 300$~K ($\beta=1$), or
320~K ($\beta=2$) at the time of the WISE observations (day~2350); a blackbody
would have cooled to 285~K.
A more detailed discussion of this will be given in a later paper.

\subsection{V476 Cyg}
This is a poorly observed nova with a resolved nebular remnant \citep{gill}.
The WISE source is within $0.\!\!''1$ of the nova position.
We have detection in WISE Bands~1 and~2 only, but no conclusions can be drawn
about the nature of the IR emission.

\subsection{V1974 Cyg}

V1974 Cyg, discovered just before maximum light by \cite{collins}. It was
shown by \cite{gehrz-cyg92a} to have strong \fion{Ne}{ii}~12.8\mic\ emission,
classifying it as a neon nova. Subsequently, V1974 Cyg developed numerous
coronal lines in the IR, including 
\fion{Al}{vi}~3.6\mic, \fion{Al}{viii}~3.7\mic, \fion{S}{ix}~1.25\mic,
\fion{Mg}{viii}~3.0\mic\ and \fion{Ca}{ix}~3.18\mic\ \citep{woodward}. 
\cite{salama-cyg} observed this CN with \iso, and reported strong
\fion{Ne}{vi}~7.6\mic, \fion{Ne}{v}~14.3\mic, \fion{Ne}{v}~24.3\mic\ emission, consistent with
its status as a neon nova; \fion{Ne}{ii}~12.8\mic\ was not detected.

The WISE SED is included in Fig.~\ref{CN1}. 
Extensive ground-based, \iso\ and \spitzer\ observations suggest that dust did
not form in the ejecta; we are therefore confident in assigning the WISE
fluxes to emission lines. Fig.~\ref{CN1} indicates that the
\fion{Ne}{iii}~15.5\mic, and \fion{O}{iv}~25.9\mic\ lines measured in 1996 with \iso\
\citep{salama-cyg} in 1996 and with \spitzer\ during
2003-2007 \citep{helton} are were likely still strong in 2010--2011.

\begin{figure}
\setlength{\unitlength}{1cm}
\begin{center}
\leavevmode
\begin{picture}(5.0,5)
\put(0.0,4.0){\includegraphics{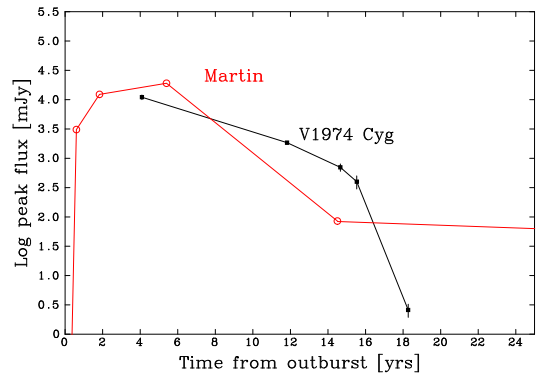}}
\end{picture}
\caption[]{Black points and line: flux
of the \fion{O}{iv}~25.9\mic\ fine-structure line in V1974~Cyg, based
on \iso, \spitzer\ and WISE data; red points and line: predicted variation of the
\fion{O}{iv} flux, calculated for DQ~Her by \cite{martin}. See text for details.
\label{v1974cyg}} 
\end{center}
\end{figure}

In Fig.~\ref{v1974cyg} we again compare the variation of the flux in the \fion{O}{iv}~25.9\mic\ line
in V1974~Cyg with that predicted by \cite{martin}, recalling the caveat that
\citeauthor{martin}'s calculations were for the specific case of DQ~Her;
we again make the assumption (see discussion of V705~Cas above) that there
is no change in the line width over the period 1992--2010 so that the peak flux
is a measure of the integrated line flux.
However we again see that the variation of the \fion{O}{iv} flux is broadly
consistent with \citeauthor{martin}'s prediction.

Although we have LC for the various Ne lines from \iso\ and \spitzer\ data
comparing these with \citeauthor{martin}'s predictions is not
appropriate as V1974~Cyg was a neon nova, while DQ~Her was not; we would not
expect agreement between theory and observation for the neon lines.

\subsection{V2361 Cyg}
\cite{russell} reported IR spectroscopy of this nova, showing a continuum
indicating dust emission at 970~K, on which were superimposed a number of broad 
($\sim2\,600$\vunit) emission lines.
Later IR observations by \cite{venturini} showed that the dust emission
had diminished, as had the extinction reported by \citeauthor{russell},
presumably due to the dispersal of the dust reported by \citeauthor{russell}.
\citeauthor{venturini} also reported the presence of the coronal lines
\fion{Si}{vi} and \fion{Ca}{viii}.

The WISE source is some $2.\!\!''5$ from the position of the nova, and
the WISE SED is included in Fig.~\ref{CN1}. In view of the cooling dust shell
and the emission lines seen in the near-IR by \cite{russell} the nature of
the emission in the WISE data is unclear.

\subsection{V2362 Cyg}
This well-observed nova had an extraordinary light curve, displaying a large amplitude 
``cusp'' that peaked some 200~days after maximum light; the cusp was superimposed on the
general decline of the light curve \citep[see][and references therein]{kimeswenger}. The 
cusp was accompanied by a rise in the X-ray flux and a shift in excitation to lower
energies \citep{lynch}. \spitzer\ observations \citep{lynch} showed that dust formed in
the ejecta at $\sim1\,400$~K some 240~days after the eruption; the dust showed emission
features at 6.37\mic, 8.05\mic\ and 11.32\mic\ that were distinct from the usual AIR
features seen in nova dust; there was also a broad ``plateau'' from 15--21\mic.
The ubiquitous \fion{O}{iv}~25.9\mic\ fine structure line was superimposed on the dust
continuum.

The WISE source is within $5''$ of the nova, just within the WISE positional uncertainty.
The SED is shown in Fig.~\ref{CN1}; in view of the presence of dust and strong \fion{O}{iv}
emission we again have insufficient information to determine the nature of the WISE emission.

\subsection{V2467 Cyg}

This nova was extensively observed in the IR. Ground-based IR observations showed 
strong coronal line emission, including 
\fion{Si}{vi}, \fion{Si}{vii}, \fion{Ca}{viii}, \fion{S}{viii} and \fion{S}{ix}.
Later the excitation had clearly increased, with 
\fion{Si}{x}, \fion{S}{xi} present \citep{perry}.
In addition to these lines, \cite{russell-cyg} reported
the rarely-seen \fion{P}{vii} feature at 1.37\mic;
the coronal lines displayed complex profiles. There was no evidence of dust \citep{mazuk},
although as noted above, dust-formation is rarely seen in ``coronal'' novae.
A possible progenitor was identified by \cite{steeghs}.
\spitzer\ IRS observations (Woodward et al., in preparation) also showed a strong emission
line spectrum, and possibly a weak dust continuum longward of $\sim15$\mic.

The WISE source is within $1.\!\!''4$ of the nova and the SED is shown in Fig.~\ref{CN2}.
It is likely that the WISE emission is due to emission lines although further analysis
(to be presented elsewhere) is needed to rule out dust emission.

\subsection{HR Del}
This nova and its stellar and nebular remnant have been very well observed and studied.
It was noted as an \iras\ source by \cite{callus}, who reported a detection in \iras\
Band~1, while \cite{HG88} reported detections in \iras\ Bands~1, 2~and~3.
It was also observed with \iso\ \citep{salama-novae}, who found that the peak flux
in the \fion{O}{iv}~25.9\mic\ fine structure line was $\sim4$~Jy in 1996 April.

The WISE SED is included in Fig.~\ref{CN2} and there is excellent positional coincidence
between the WISE source and the nova. Although it was a very slow nova
($t_3=231$~day) HR Del seems not to have been a dust-former. It seems certain
therefore that the WISE fluxes are due to line emission, and we note that
\fion{O}{iv} seems still to have been strong in 2010. 

\subsection{DM Gem}
This is a poorly-studied nova. The WISE source is within
$2''$ of the nova, and there are secure detections in Bands~1 and~2.
However there is insufficient information to draw any conclusions as to
the origin of the WISE emission.

\subsection{DN Gem}

Another poorly-studied nova, albeit with a known orbital period (3.068~hours)
and a resolved nebular remnant \citep[see][]{CNII}. The WISE source is within
$0.\!\!''5$ of the nova; there are detections in Bands~1 and~2, although neither
is strong. Again we have insufficient information to draw any conclusions as to
the origin of the WISE emission.

\subsection{DQ Her}
This is a very-well observed and studied nova, which is an eclipsing binary.
The deep minimum in the 1934 light curve was interpreted by \cite{McL} as being
due to dust-formation: DQ~Her is the proto-type of dust-forming novae. It has a
resolved and well-studied nebular remnant \citep[see][and references therein]{OBBode},
the optical spectrum of which has been studied for well over 50~years. \cite{martin}
has used {\sc cloudy} \citep{cloudy} to model the nebula over
the period 1934--1982, and has predicted the evolution of the line emission,
from the ultra\-violet to the far-IR.

The WISE source is within $0.\!\!''1$ of the nova and the SED is shown in Fig.~\ref{CN2}.
The WISE Band 1--3 data are reasonably well fitted by a 3\,310K blackbody, corresponding
to an early M dwarf star at 485~pc \citep{martin}. It is therefore likely that the WISE data are
detecting emission from the stellar photosphere.

\subsection{V533 Her}
This nova has a resoved remnant \citep[see][and references therein]{OBBode}.
It is also unusual in that its LC is known prior to its 1963 outburst
\citep[see][]{warner-CNII}.

The WISE source is close ($0.\!\!''2$) to the nova but there is insufficient
information to draw any conclusion about the nature of the emission.


\subsection{GK Per}
The first bright nova of the 20$^{\rm th}$ century, this is a very well-studied object.
It has an evolved (K2IV) secondary \citep{darnley}, a resolved nebular remnant
\citep{OBBode,liimets}, and seems to lie within a planetary nebula
\citep{GKPer}. It was not detected in the \iras\ survey \citep{HG88}.

The WISE source is some $3''$ from the nova, within the WISE positional uncertainties;
the WISE SED is shown in Fig.~\ref{CN2}. The data in all four WISE bands seem consistent
with a 3\,390~K blackbody, corresponding to an early K2 subgiant.

\subsection{RR Pic}
RR Pic has a resolved nebular remnant that has been studied over a range of wavelenths
\citep[see][and references therein]{OBBode}.
RR~Pic was not detected by \iras\ \citep{HG88}.

The WISE soure is less than $0.\!\!''2$ from the position of the nova; the WISE SED is shown
in Fig.~\ref{CN2}. The Band~1 and~2 data can be fit with a $\sim2\,230$~K blackbody, although the 
temperature is very poorly constrained.

\subsection{V1186 Sco}

V1186~Sco was a well studied nova. Its LC displayed a slow rise to maximum, and there
was a prominent secondary maximum \citep[see][]{schwarz}. It was observed by \spitzer\
\citep{schwarz} and, despite that fact that it was a CO nova, there was no evidence
of dust emission.

The WISE source is $<1''$ from the position of the nova. The WISE SED is shown in
Fig.~\ref{CN2}. It is superficially similar to a $\sim500$~K blackbody but as V1186~Sco
is known not to have been a dust-producer, we can rule out dust emission. The emission
detected by WISE is very likely due to line emission.

\subsection{V1280 Sco}
V1280~Sco was a dusty nova with a complex visual LC and seemingly invariant dust
temperature \citep[see][]{chesneau2}; high spatial resolution observations by
\cite{chesneau} showed that the dust resided in polar caps.
The WISE source is $\sim2.\!\!''8$ from the nova, well within the WISE positional
uncertainties, and the SED is shown in Fig.~\ref{CN2}. The nova is strongly detected in
all four WISE Bands and the data are consistent with a 510~K blackbody.

\subsection{CT Ser}
This is a poorly-studied nova. The WISE source is within $0.\!\!''1$ of the
position of the nova. The nova is detected in Bands~1 and~2 only and the data 
are insufficient to draw any conclusions about the nature of the emission.

\subsection{V382 Vel}

V382 Vel was a fast nova; it is in the class of ONe novae, based on the detection of prominent
\fion{Ne}{ii}12.8\mic\ emission at 43.6 days after maximum
\citep{woodward-vel}. \spitzer\ spectra of V382 Vel \citep{helton} showed
strong emission by \fion{Ne}{ii}12.8\mic, \fion{Ne}{iii}15.5\mic\ and
\fion{O}{iv}25.9\mic. Emission lines of \fion{Ne}{v} were present at 14.3\mic\
and 24.30\mic, but at a much lower level than the other neon species; no
\fion{Ne}{vi}7.6\mic\ emission was detected.  There was also weak emission from
\fion{S}{iv}10.5\mic\ and \fion{Ar}{iii}8.9\mic. In keeping with its neon nova status,
V382~Vel did not produce any dust.

The WISE source is $\sim0.\!\!''5$ from the nova and is strongly detected in all four
WISE Bands. The SED is shown in Fig.~\ref{CN2}; in view of the fact that V382~Vel was
not a dust-fomer
the WISE data indicate that the lines seen by \spitzer\ were still strong in 2010. 


\onecolumn

\setlongtables
\LTcapwidth=8.4in
\small
\begin{landscape}

\begin{table*}
\begin{center}
\caption{WISE detections of post-2011 novae. \label{pre_fluxes}}  
\begin{tabular}{lrllccccccc}  \hline
Nova & \multicolumn{1}{c}{Outburst} & \multicolumn{2}{c}{RA, Dec (J2000)}    & Date of first    & B1 (3.3\mic) & B2 (4.6\mic) & B3 (12\mic) &  B4 (22\mic) & $\Delta\theta$ &Comment \\
     & \multicolumn{1}{c}{Date UT}     & \multicolumn{2}{c}{Upper value: WISE}  & WISE Observation & (mJy)       &   (mJy)      & (mJy)       &   (mJy)    &  &         \\
     & \multicolumn{1}{c}{\sc yyyy-mm-dd.dd}         &  \multicolumn{2}{c}{Lower value: GCVS} & YYYY-MM-DD       & ({\bf Jy})  &  ({\bf Jy}) & ({\bf Jy})   & ({\bf Jy})  &   (arcsec)      \\ \hline

PNV J14250600--5845360 & 2012-04-5.47    & \RA{14}{25}{06~~}  & \dec{--58}{45}{36~~} & 2010-02-13 & $2.22\pm0.07$	& $1.03\pm0.04$ & $<0.37$ & $<2.26$  & 2.12 &\\
Nova Cen 2012b  &                  & \RA{14}{25}{06.10}  & \dec{--58}{45}{34.07} &  & &  &  &   &  &\\
               &     	          &  &   & & & & & & &  \\
V959 Mon  & 2012-08-9.8 & \RA{06}{39}{38.57}  & \dec{+05}{53}{51.59} & 2010-03-23 & $0.41\pm0.01$	& $0.19\pm0.02$ & $<0.610$ & $<2.668$  & 2.39 &\\
               &                  & \RA{06}{39}{38.74}  & \dec{+05}{53}{52.0} &  & &  &  &   &  &\\
               &     	          &  &   & & & & & & &  \\
V2677 Oph   & 2012-05-19.48  & \RA{17}{39}{56~~}  & \dec{--24}{47}{42~~} & 2010-03-14 &$31.12\pm0.75$ & $13.75\pm0.35$ & $4.04\pm0.26$ & $3.95\pm1.02$ & 5.31\\
Nova Oph 2012b               &                  & \RA{17}{39}{55.84}  & \dec{--24}{47}{37.21}  & & & & \\
               &     	          &  &   & & & & & & &  \\
V5590 Sgr & 2012-04-23.7  & \RA{18}{11}{03.71}   & \dec{--27}{17}{29.37} &2010-03-21& {\bf 1.17$\pm$0.11} & {\bf 1.30$\pm$0.05} & {\bf
0.54$\pm$0.01} & {\bf 0.29$\pm$0.01} & 1.58 & \iras, 2MASS, \\
Nova Sgr 2012b   &     &  \RA{18}{11}{03.75} &  \dec{--27}{17}{28.4} & & & & & & & ARAKI source \\
              &     	          &   &   & & & & & & &     \\
               &     	          &  &   & & & & & & &  \\

\hline\hline 
\end{tabular}
\end{center}
\end{table*}

\begin{table*}
 \centering
  \caption{Post-2011 novae not detected by {\it WISE}. \label{non-det2}}
  \begin{tabular}{lll}\hline 
GCVS name & Other name & Outburst/discovery date (UT) \\ \hline
 &Nova Aql 2012 & 2012 Oct 20.4 \\
V834 Car &Nova Car 2012 & 2012 Feb 26.5\\
V1368 Cen & Nova Cen 2012 & 2012 Mar 23.4 \\
V1369 Cen & Nova Cen 2013 & 2013 Dec 2.7 \\
V339 Del &Nova Del 2013 & 2013 Aug  14.6\\
PR Lup &Nova Lup 2011 & 2011 Aug 4.7 \\
V2676 Oph & Nova Oph 2012a  & 2012-03-25.79 \\
V2677 Oph &Nova Oph 2012b & 2012 May 19.5 \\
V1311 Sco &Nova Sco 2010 No.2 & 2010 Apr 25.8\\
V1312 Sco &Nova Sco 2011 & 2011 Jun 1.6 \\
V1313 Sco &Nova Sco 2011 No. 2 & 2011 Sep 06.4 \\
V1324 Sco &Nova Sco 2012 & 2012 May 22.8 \\
V5588 Sgr & Nova Sgr 2011b & 2011 Mar 27.8 \\
V5589 Sgr & Nova Sgr 2012 & 2012 Apr 21.0 \\
V5591 Sgr & Nova Sgr 2012c & 2012 Jun 26.6 \\
V5592 Sgr & Nova Sgr 2012d &  2012 July 7.5 \\
V5593 Sgr & Nova Sgr 2012e & 2012 Jul 16.5 \\
 \hline\hline
\end{tabular}
\end{table*}

\end{landscape}

\twocolumn

\section{Post-WISE novae observed by WISE}

A number of novae erupted after WISE completed its mission. This gives us an opportunity
of obtaining a glimpse of the mid-IR properties of the nova progenitors. At the time of writing,
WISE detected four of these; the data are listed in Table~\ref{pre_fluxes}. However care
should be exercised in using and interpreting these data as they stand as the WISE catalogue
indicates
that there may have been significant flux variations during the course of the WISE mission.
We therefore defer a detailed discussion of these objects to a later paper and give only
a brief description here.

\subsection{V959 Mon}

V959 Mon (Nova Mon 2012) is exceptional in that it was a $\gamma$-ray source \citep{mon12}.
It is a weak source in WISE Bands 1 and 2, and there is some evidence for variability.


\subsection{V5590 Sgr}
This slow  nova (still referred to in the literature as Nova Sagittarii 2012b),
was discovered on 2012 April 23, although it seems to have been
bright at least a year earlier \citep{SMARTS}; it has been a SMARTS target for optical
and IR photometry, and spectroscopy \citep{SMARTS}.

\begin{figure}
\setlength{\unitlength}{1cm}
\begin{center}
\leavevmode
\begin{picture}(5.0,5)
\put(0.0,4.0){\includegraphics{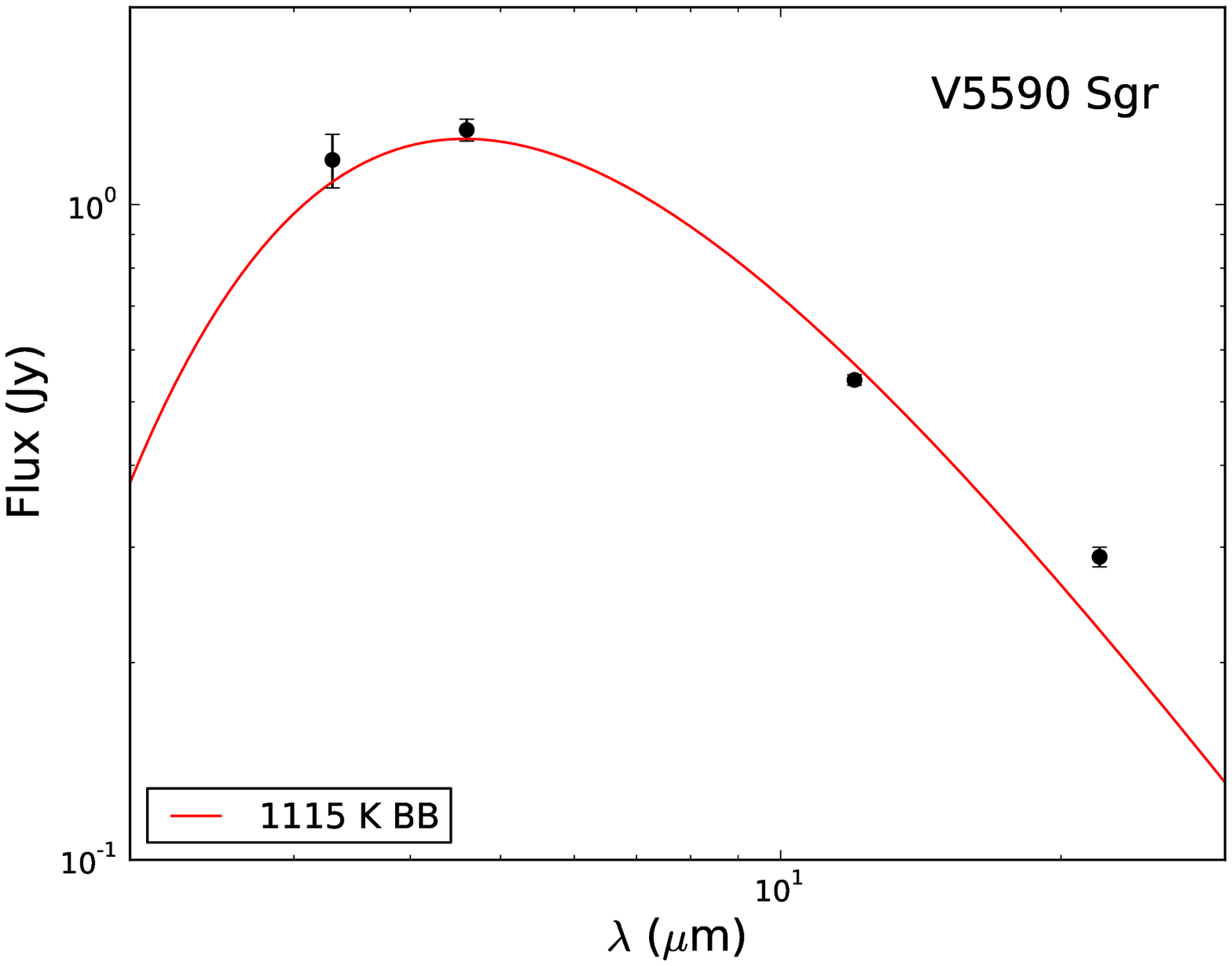}}
\end{picture}
\caption[]{WISE SED of V5590 Sgr. 
Errors are smaller than plotted points if not shown.
\label{post-wise}} 
\end{center}
\end{figure}

The WISE SED is shown in Fig.~\ref{post-wise}. The data are consistent
with a cool continuum, at $\sim1\,115$~K. While it is tempting to associate this emission
with hot dust (possibly dust from a giant companion wind)
it is not possible to rule out a V838~Mon-type event \citep[see e.g.][]{evans_v838}.

\subsection{Non-detections}
We list in Table~\ref{non-det2} post-2011 novae that were not detected in the
WISE survey. In some cases the  
novae listed have WISE sources close to the position of the nova but not within
the positional uncertainties of the WISE survey.

\section{Conclusions} 
We have searched the WISE database for IR sources that might be associated with mature novae
(including classicals and recurrents). We have found 36 possible associations with old
novae; these show a mix of photospheric, line and dust emission, or a combination of two
of these; forty-eight novae (including those which erupted during or after the WISE mission)
seem to have no counterparts in the WISE survey.

We have also searched the WISE database for sources associated with novae that erupted
after the WISE mission had terminated, giving some information about nova progenitors.

In future papers we will be reporting on detailed modelling of the WISE data.

\section*{Acknowledgments}

We thank Dr. Jim Lyke for helpful comments on an earlier version of this paper,
and Alex d'Angelo for making an initial trawl for novae in the WISE archive.

RDG was supported by NASA and the United States Air Force.

This publication makes use of data products from the Wide-field Infrared Survey
Explorer, which is a joint project of the University of California, Los Angeles,
and the Jet Propulsion Laboratory/California Institute of Technology, funded by
the National Aeronautics and Space Administration.

This research has made use of the SIMBAD database, operated at CDS, Strasbourg,
France.

\bsp

\label{lastpage}

\end{document}